\documentclass[aps,prbl,twocolumn,10pt,longbibliography]{revtex4-2}
\usepackage[utf8]{inputenc}
\usepackage{libertine}
\usepackage[libertine]{newtxmath}
\usepackage{mathtools,graphicx,microtype,siunitx}
\usepackage[cal=boondoxo,frak=boondox]{mathalfa}
\usepackage[colorlinks=true,allcolors=blue]{hyperref}

\begin{document}
\title{Two-body problem for two-dimensional electrons in Bernervig-Hughes-Zhang model}

\author{Vladimir~A.~Sablikov}

\affiliation{Kotelnikov Institute of Radio Engineering and Electronics, Russian Academy of Sciences,
Fryazino, Moscow District, 141190, Russia}

\begin{abstract}
We study the two-body problem for two-dimensional electron systems in a symmetrized Bernevig-Hughes-Zhang model which is widely used to describe topological and conventional insulators. The main result is that two interacting electrons can form bound states with the energy in the gap of the band spectrum. The pairing mechanism can be interpreted as the formation of a negative reduced effective mass of two electrons. The problem is complicated because the relative motion of the electrons is coupled to the center-of-mass motion. We consider the case of zero total momentum. Detail calculations are carried out for the repulsive interaction potential of steplike form. The states are classified according to their spin structure and two-particle basis functions that form a given bound state. We analyze the spectra and electronic structure of the bound states in the case of both topological and trivial phases and especially focus on effects originating from the band inversion and the coupling of the electron and hole bands. In the trivial phase and the topological phase with the large coupling parameter $a$, the bound state spectra are qualitatively similar. However, when $a$ is less a certain value, the situation changes dramatically. In the topological phase, new states arise  with a higher binding energy at lower interaction potential, which evidences that the band inversion can favor pairing the electrons.
\end{abstract}
\maketitle

\section{Introduction}
\label{Intro}
The electron-electron interactions in topological insulators (TIs) currently are one of the most challenging problems in which one can expect the emergence of new and nontrivial properties of electronic systems. The role of the electron-electron interactions in TIs is still poorly understood, but it is already clear that a lot of highly unusual effects arises due to the interplay between Coulomb interactions and topological aspects of matter~\cite{QiChangRevModPhys2011}. Recent experiments clearly demonstrate that in many cases the inclusion of the electron-electron interaction is crucially important for understanding the electron transport in TIs (see, e.g., Refs.~\cite{WangPRB2011,LiWangFuPRL2015}). The theoretical researches are focused mainly on the influence of the electron-electron interactions on the topological phase transitions and on the possibility of a topological phase to be formed due to the interaction (for a review of recent studies in this field on the two-dimensional (2D) TIs see, e.g., Ref.~\cite{HohenadlerJPCM2013}).

There is another aspect of the many-body problem associated with the formation of stable or metastable complexes of two or more electrons, such as Cooper pairs, excitons, exciton complexes, etc. Investigations in this direction are carried out very intensively for conventional materials for many years and currently continue to attract great interest~\cite{combescot2015excitons}. However, such states are still insufficiently studied for topologically nontrivial electronic systems, although it is clear that new properties of the bound states can appear thanks to the unusual band structure of TIs. 

Recent studies were focused on electron-hole bound states. It was found that chiral excitons arise on a surface of TIs with a magnetically induced gap in the surface state spectrum~\cite{GaratePRB2011}. They differ from conventional excitons by a chiral structure and a modified spectrum. Chiral excitons give rise to resonant manifestations in Faraday and Kerr effects~\cite{EfimkinPRB2013}. Similarly, edge excitons were found to be formed in 2D TIs in the presence of an in-plain magnetic field~\cite{entin2016edge}. In the bulk of the crystal, the exciton states are affected by the geometrical properties of the Bloch bands. The Berry curvature and quantum geometric tensor essentially modify the exciton states and their spectrum~\cite{ZhouPRL2015,SrivastavaPRL2015}. The studies of excitons in topologically trivial narrow-gap materials revealed a substantial dependence of the exciton properties on the electron dispersion in the bands. Such investigations were carried out in recent years for the quasirelativistic dispersion of electrons and holes in a gapped graphene and carbon nanotubes~\cite{IyengarPRB2007,HartmannPRB2011,BermanPRA2013,HartmannAIP2016}.

In the present paper, we address to the problem of two interacting electrons. In this case, the formation of bond states does not seem obvious because of the Coulomb repulsion. Nevertheless, bound states can be formed because of peculiarities of the band structure. This conclusion can be drawn from recent studies of two-electron states in graphene. Sabio, Sols, and Guinea~\cite{SabioPRB2010} investigated the problem of two interacting Dirac fermions and revealed a singular behavior of the two-particle wave function at a definite distance between the particles, which evidenced a partial localization of the interacting electrons. The existence of quasilocalized two-particle states in this case was demonstrated in Ref.~\cite{LeePRB2012}. Further studies showed that the formation of stationary two-particle states with localized wave function becomes possible when the quasiparticle dispersion substantially deviates from the linear one. First, Mahmoodian and Entin found that the trigonal warping of the spectrum results in the formation of the excitonlike states in some regions of the momentum space~\cite{MahmoodianEntinEPL2013}. Then, Marnham and Shytov introduced quadratic momentum terms into the single-particle kinetic energy and came to the conclusion that the bound, Cooper-pairlike, states could appear in double-layered structures, but they were metastable~\cite{MarnhamShytovPRB2015}.

The situation of TIs is obviously more interesting, because in this case there is a gap in the single-particle spectrum and the single-particle states have a more complex orbital structure. The electronic states are a superposition of the states of the electron and hole bands, which are characterized by effective masses of opposite sign. One can therefore expect a nontrivial dynamics of the particles under the action of the Coulomb forces since the relative motion of the particles is determined by a reduced effective mass, the sign of which is not obvious in advance, i.e., without knowledge of the orbital composition of the two-particle wave function which in its turn is determined by the solution of corresponding Schr\"odinger equation.

This conjecture is supported by the results of recent studies of the bound states localized at impurities with a short-range potential~\cite{SablikovPSSR2014,SablikovPRB2015,SlagerPRB2015}. It turns out that the potential of any sign produces bound states of two kinds in the energy gap of the 2D TIs, in contrast to the topologically trivial case where only one bound state exists. For example, in the case of an impurity with negative potential, one state is formed as a result of the attraction of the electronlike quasiparticle. The captured particle is localized in the center. Other state, on the contrary, arises as a result of the repulsion of the holelike quasiparticle. In this state, the particle is localized around the impurity similarly to edge states. In other words, an impurity produces a bound state in both cases: when the impurity attracts a particle or repels it.

In this connection a natural question arises whether two electrons form a bound state in 2D TIs when the Coulomb force acts between them? To answer this question in the present work we study two-particle states within the model proposed by Bernevig, Hughes, and Zhang~\cite{BHZScience2006} (BHZ). The model is widely used for 2D TIs, but it describes also a trivial phase under appropriately chosen parameters, so that we can compare the results obtained in both cases to reveal effects that arise only in a topological phase. Two-electron states have not yet been studied in the BHZ model. We solve this problem and show that bound states indeed arise. It is found that the bound states can be formed in both the topological and trivial phases at any sign of the pair interaction potential. However, in the topological phase, new bound states appear in addition to those in the trivial phase. They arise at lower interaction potential and have a higher bounding energy. We study general properties of the bound state spectra and classify the states according to their spin structure as singletlike and tripletlike ones. 
With respect to orbital degrees of freedom, the bound states are well classified only in the case of small interaction potential where the states are separated into two groups. It the one group, the states are mainly formed by basis states in which both electrons are in the same (electron or hole) band. It the other group, the pairing electrons are in the different bands.

The paper is organized as follows. Section~\ref{general} presents general equations. Here, we also classify the two-particle states and simplify the problem by addressing to the case of zero center-of-mass momentum and to a model potential. In Sec.~\ref{qualitative}, we present qualitative arguments explaining the bound state formation. Section~\ref{singlet} is devoted to singletlike bound states. The tripletlike states are considered in Sec.~\ref{triplet}. In Sec.~\ref{invert}, a specific case is studied to show that new bound states arise due to the band inversion in the topological phase. In Sec.~\ref{trivial} the topologically trivial case is considered and the band-inversion effect is discussed. Finally, in Sec.~\ref{conclude}, we summarize the results.

\section{General equations}
\label{general}
We start with a statement of the two-body problem in the BHZ model. The BHZ model presents single-particle electronic states in the frame of the $\mathbf{kp}$ theory with using four-band basis $(|E\uparrow\rangle,|H\uparrow\rangle,|E\downarrow\rangle,|H\downarrow\rangle)^T$, where $|E\uparrow\rangle$ and $|E\downarrow\rangle$ are a superposition of the electron- and light-hole states with the moment projection $m_J=\pm 1/2$; $|H\uparrow\rangle$ and $|H\downarrow\rangle$ are the heavy-hole states with $m_J=\pm 3/2$. The single-particle Hamiltonian that determines the spinor of the envelope functions reads 
\begin{equation}\label{H_0}
 \hat{H}_0(\mathbf{\hat{k}})=
\begin{pmatrix}
 \hat{h}(\mathbf{\hat{k}}) & 0\\
 0 & \hat{h}^*(-\mathbf{\hat{k}})
\end{pmatrix}
\end{equation}
\begin{equation}\label{h(k)}
\hat{h}(\mathbf{\hat{k}})=
\begin{pmatrix}
 M\!-\!B\hat{k}^2 & A(\hat{k}_x\!+\!i\hat{k}_y)\\
 A(\hat{k}_x\!-\!i\hat{k}_y) & -M\!+\!B\hat{k}^2
\end{pmatrix}\,,
\end{equation} 
where $\mathbf{\hat{k}}$ is the quasimomentum operator, $A$, $B$, and $M$ are the parameters of the BHZ model. Here, for simplicity, we do not take into account the spin-orbit interaction, which can actually be present due to structural inversion asymmetry and bulk inversion asymmetry. The terms describing the asymmetry of the electron and hole bands also are dropped for simplicity. These assumptions do not have a decisive impact on the results but greatly simplify the calculations. The BHZ model describes both topological and trivial phases of a 2D electron system in a crystal. Trivial phase is realized at the ordinary arrangement of the electron and hole bands, when $M/B<0$. In the topological phase the band structure is inverted, $M/B>0$. 

Two-particle wave functions are represented by a spinor of 16th order, $\Psi(\mathbf{r}_1,\mathbf{r}_2)=\left(\psi_1,\psi_2,\psi_3, ...,\psi_{16}\right)^T$, which defines the envelope functions in the basis:
\begin{multline}\label{basis_f}
 \left(|E\uparrow, E\uparrow\rangle,|E\uparrow, H\uparrow\rangle,|E\uparrow, E\downarrow\rangle,|E\uparrow, H\downarrow\rangle,\right.\\ |E\downarrow, H\uparrow\rangle,\dots, \left.|H\downarrow, E\downarrow\rangle,|H\downarrow, H\downarrow\rangle\right)^T.
\end{multline} 

The Hamiltonian of two interacting electrons has the form
\begin{equation}
 \hat{H}(1,2)=\hat{H}_0(\mathbf{\hat{k}}_1)\oplus\hat{H}_0(\mathbf{\hat{k}}_2)+V(\mathbf{r}_1-\mathbf{r}_2)\cdot\hat{\mathbf{I}}_{16\times 16}\,,
\end{equation} 
where $V(\mathbf{r})$ is the pair interaction potential which is supposed to be a given function.

The wave function $\Psi(\mathbf{r}_1,\mathbf{r}_2)$ is determined by the Schr\"odinger equation
\begin{equation}\label{Schrodinger}
 \hat{H}(1,2)\Psi(\mathbf{r}_1,\mathbf{r}_2)=E\Psi(\mathbf{r}_1,\mathbf{r}_2)\,.
\end{equation} 

Due to the block-diagonal structure of the single-particle Hamiltonian~(\ref{H_0}), the Schr\"odinger equation~(\ref{Schrodinger}) splits into four uncoupled equations for the following wave functions:
\begin{equation}\label{4_Psi}
 \begin{split}
 &\Psi_1(1,2)\!=\!
 \begin{pmatrix}
  \psi_1\!\cdot\!|E\!\uparrow E\!\uparrow \rangle\\ \psi_2\!\cdot\!|E\!\uparrow H\!\uparrow \rangle\\ \psi_5\!\cdot\!|H\!\uparrow E\!\uparrow \rangle \\ \psi_6\!\cdot\!|H\!\uparrow H\!\uparrow \rangle
 \end{pmatrix}\!,\,
 \Psi_2(1,2)\!=\!
 \begin{pmatrix}
  \psi_3\!\cdot\!|E\!\uparrow E\!\downarrow \rangle\\ \psi_4\!\cdot\!|E\!\uparrow H\!\downarrow \rangle\\ \psi_7\!\cdot\!|H\!\uparrow E\!\downarrow \rangle\\ \psi_8\!\cdot\!|H\!\uparrow H\!\downarrow \rangle 
 \end{pmatrix}\!,\,\\
 &\Psi_3(1,2)\!=\!
 \begin{pmatrix}
  \psi_9\!\cdot\!|E\!\downarrow E\!\uparrow \rangle \\ \psi_{10}\!\cdot\!|E\!\downarrow H\!\uparrow \rangle \\ \psi_{13}\!\cdot\!|H\!\downarrow E\!\uparrow \rangle \\ \psi_{14}\!\cdot\!|H\!\downarrow H\!\uparrow \rangle
 \end{pmatrix}\!,\,
 \Psi_4(1,2)\!=\!
 \begin{pmatrix}
  \psi_{11}\!\cdot\!|E\downarrow E\!\downarrow \rangle \\ \psi_{12}\!\cdot\!|E\!\downarrow H\!\downarrow \rangle \\ \psi_{15}\!\cdot\!|H\!\downarrow E\!\downarrow \rangle \\ \psi_{16}\!\cdot\!|H\!\downarrow H\!\downarrow \rangle
 \end{pmatrix}\!.
 \end{split}
\end{equation}  
Here, for clarity, we have written both the envelope functions $\psi_1, \psi_2, \dots$ and the corresponding basis functions. The states described by $\Psi_1(1,2)$ and $\Psi_4(1,2)$ are composed of the spin-up and spin-down orbitals. Therefore, they can be conventionally classified as tripletlike states. Similarly, the states $\Psi_2(1,2)$ and $\Psi_3(1,2)$ can be called singletlike ones. These terms are not strict here. In Sec.~\ref{singlet} it will be shown that the wave functions $\Psi_2(1,2)$ and $\Psi_3(1,2)$ describe the same bound state.

The wave functions $\Psi_j(1,2)$ are determined by equations of the following form:
\begin{equation}
 \left\{\hat{H}_j-\bigl[\varepsilon-2v(\mathbf{r})\bigr] \mathbf{I}_{4\times 4} \right\} \Psi_j(\mathbf{r},\mathbf{R})=0.
\end{equation} 
Here and in what follows we use dimensionless notations:
\begin{equation}\label{dim_less_param}
 \varepsilon\!=\!\frac{E}{|M|},\; \mathbf{r'}\!=\!\mathbf{r}\sqrt{\frac{|M|}{|B|}},\; a\!=\!\frac{A}{\sqrt{|MB|}},\; v(\mathbf{r'})=\frac{V(\mathbf{r})}{2|M|}\,.
\end{equation} 
For convenience, the prime in the variable $\mathbf{r'}$ will be omitted. To separate the topological and trivial phases we introduce a parameter $\lambda=M/|M|=\pm 1$ and assume that $B<0$. In this case, $\lambda=1$ corresponds to the trivial phase and $\lambda=-1$ corresponds to the topological phase.

The operators $\hat{H}_j$ are $4\times 4$ matrices, the elements of which are expressed via the operators $\hat{\mathbf{k}}_1$ and $\hat{\mathbf{k}}_2$. Before we present the equations of motion in an explicit form, it is meaningful to modify the wave functions taking into account that the system is translationally invariant. 

In order to study the bound states it would be natural to try to separate the relative motion of the particles from their movement as a whole. Therefore we switch to the center-of-mass frame, defining the new coordinates: $\mathbf{R}=(\mathbf{r}_1+\mathbf{r}_2)/2$ and $\mathbf{r}=\mathbf{r}_1-\mathbf{r}_2$. However, within the BHZ model the relative motion and the motion of the center of mass are not separated because the nondiagonal terms in Eq.~(\ref{h(k)}), which determine the coupling of the electron and hole bands, depend on the momenta of each particle. Nevertheless, since the system is translationally invariant, the wave function can be represented in the form: 
\begin{equation}
 \Psi_j(\mathbf{R},\mathbf{r})=\Psi_{j,\mathbf{K}}(\mathbf{r})e^{i\mathbf{KR}},\,
\end{equation} 
where $\mathbf{K}$ is the total momentum of the pair. 

Of most interest are the functions $\Psi_{j,\mathbf{K}}(\mathbf{r})$ that determine the relative motion. The fact that $\Psi_{j,\mathbf{K}}(\mathbf{r})$ is explicitly dependent on $\mathbf{K}$ means that the spectrum of the bound states and their structure depend on the total momentum. 

The equations defining the functions $\Psi_{j,\mathbf{K}}(\mathbf{r})$ are different for all four states given by Eq.~\ref{4_Psi}. In the case of the tripletlike state with the moments up, the function $\Psi_{1,\mathbf{K}}(\mathbf{r})=\left(\psi_1,\psi_2,\psi_5,\psi_6\right)^T$ is determined by the following equation system:
\begin{widetext}
\begin{equation}\label{triplet_gen}
\left\{
\begin{array}{rl}
 \left(-\dfrac{\varepsilon}{2}+\lambda+\hat{k}^2+\dfrac{K^2}{4}+v(r)\right)\psi_1(\mathbf{r})-\dfrac{a}{2}\left(\hat{k}_+-\dfrac{K_+}{2}\right)\psi_2(\mathbf{r})+\dfrac{a}{2}\left(\hat{k}_++\dfrac{K_+}{2}\right)\psi_5(\mathbf{r}) &=0\\
 -\dfrac{a}{2}\left(\hat{k}_--\dfrac{K_-}{2}\right)\psi_1(\mathbf{r})+\left(-\dfrac{\varepsilon}{2}+\mathbf{\hat{k}K}+v(r)\right)\psi_2(\mathbf{r})+\dfrac{a}{2}\left(\hat{k}_++\dfrac{K_+}{2}\right)\psi_6(\mathbf{r}) &=0\\
 \dfrac{a}{2}\left(\hat{k}_-+\dfrac{K_-}{2}\right)\psi_1(\mathbf{r})+\left(-\dfrac{\varepsilon}{2}-\mathbf{\hat{k}K}+v(r)\right)\psi_5(\mathbf{r})-\dfrac{a}{2}\left(\hat{k}_+-\dfrac{K_+}{2}\right)\psi_6(\mathbf{r}) &=0\\
 \dfrac{a}{2}\left(\hat{k}_-+\dfrac{K_-}{2}\right)\psi_2(\mathbf{r})-\dfrac{a}{2}\left(\hat{k}_--\dfrac{K_-}{2}\right)\psi_5(\mathbf{r}) + \left(-\dfrac{\varepsilon}{2}-\lambda-\hat{k}^2-\dfrac{K^2}{4}+v(r)\right)\psi_6(\mathbf{r}) &=0\,.
\end{array}\right. 
\end{equation}
The wave function $\Psi_{2,\mathbf{K}}(\mathbf{r})=\left(\psi_3,\psi_4,\psi_7,\psi_8\right)^T$ of the singletlike state is defined by the following equations:
\begin{equation}\label{singlet_gen}
\left\{
\begin{array}{rl}
 \left(-\dfrac{\varepsilon}{2}+\lambda+\hat{k}^2+\dfrac{K^2}{4}+v(r)\right)\psi_3(\mathbf{r})+\dfrac{a}{2}\left(\hat{k}_--\dfrac{K_-}{2}\right)\psi_4(\mathbf{r})+\dfrac{a}{2}\left(\hat{k}_++\dfrac{K_+}{2}\right)\psi_7(\mathbf{r}) &=0\\
 \dfrac{a}{2}\left(\hat{k}_+-\dfrac{K_+}{2}\right)\psi_3(\mathbf{r})+\left(-\dfrac{\varepsilon}{2}+\mathbf{\hat{k}K}+v(r)\right)\psi_4(\mathbf{r})+\dfrac{a}{2}\left(\hat{k}_++\dfrac{K_+}{2}\right)\psi_8(\mathbf{r}) &=0\\
 \dfrac{a}{2}\left(\hat{k}_-+\dfrac{K_-}{2}\right)\psi_3(\mathbf{r})+\left(-\dfrac{\varepsilon}{2}-\mathbf{\hat{k}K}+v(r)\right)\psi_7(\mathbf{r})+\dfrac{a}{2}\left(\hat{k}_--\dfrac{K_-}{2}\right)\psi_8(\mathbf{r}) &=0\\
 \dfrac{a}{2}\left(\hat{k}_-+\dfrac{K_-}{2}\right)\psi_4(\mathbf{r})+\dfrac{a}{2}\left(\hat{k}_+-\dfrac{K_+}{2}\right)\psi_7(\mathbf{r}) + \left(-\dfrac{\varepsilon}{2}-\lambda-\hat{k}^2-\dfrac{K^2}{4}+v(r)\right)\psi_8(\mathbf{r}) &=0\,.
\end{array}\right. 
\end{equation}
\end{widetext}
The wave functions $\Psi_{4,\mathbf{K}}(\mathbf{r})$ and $\Psi_{3,\mathbf{K}}(\mathbf{r})$ are described by similar equations which are not presented here for the sake of brevity.

In Eqs.~(\ref{triplet_gen}) and (\ref{singlet_gen}), $\hat{\mathbf{k}}=(\hat{\mathbf{k}}_1-\hat{\mathbf{k}}_2)/2$ is the operator of the relative momentum, $\hat{k}_{\pm}=\hat{k}_x \pm i\hat{k}_y$ and $K_{\pm}=K_x \pm iK_y$. 

\subsection{The case of zero total momentum}
\label{center-of-mass}

In order to understand the essential features of the problem, of most interest is the case of zero total center-of-mass momentum. One can expect that in this case the pairing effect is particularly important since the kinetic energy of the pair is minimal. This is also the simplest case for calculations since Eqs~(\ref{triplet_gen}) and (\ref{singlet_gen}) are considerably simplified. From these equations, one can suppose that the finite value of the total momentum does not strongly change the solutions as long as $K$ is small enough. This is why we will focus on the case where $\mathbf{K}=0$.

To be specific, consider first the state described by the wave function $\Psi_{2,\mathbf{K}}(\mathbf{r})$. In the case of $\mathbf{K}=0$, Eq.~(\ref{singlet_gen}) is simplified as follows:
\begin{equation}\label{singlet_K0}
\left\{
\begin{array}{rl}
 \left[2v(r)\!-\!\varepsilon\!+\!2\lambda\!+\!2\hat{k}^2\right]\psi_3(\mathbf{r})\!+\!a\hat{k}_-\psi_4(\mathbf{r})\!+\!a\hat{k}_+\psi_7(\mathbf{r}) &\!=0\\
 a\hat{k}_+\psi_3(\mathbf{r})+\left[2v(r)-\varepsilon\right]\psi_4(\mathbf{r})+a\hat{k}_+\psi_8(\mathbf{r}) &\!=0\\
 a\hat{k}_-\psi_3(\mathbf{r})+\left[2v(r)-\varepsilon\right]\psi_7(\mathbf{r})+a\hat{k}_-\psi_8(\mathbf{r}) &\!=0\\
 a\hat{k}_-\psi_4(\mathbf{r})\!+\!a\hat{k}_+\psi_7(\mathbf{r})\!+\!\left[2v(r)\!-\!\varepsilon\!-\!2\lambda\!-\!2\hat{k}^2\right]\psi_8(\mathbf{r}) &\!=0\,.
\end{array}\right. 
\end{equation}

For further analysis, it is convenient to go to polar coordinates ($r,\varphi$) and expand the wave functions in the Fourier series:
\begin{equation}\label{Psi2_F}
 \Psi_{2,K=0}(\mathbf{r})\!=\!\sum\limits_m \Psi_{2m} e^{im\varphi}\!=\!\sum\limits_m
 \begin{pmatrix}
 \psi_{3m}(r) \\ \psi_{4m}(r) e^{i\varphi} \\ \psi_{7m}(r) e^{-i\varphi} \\ \psi_{8m}(r)
 \end{pmatrix}
 e^{im\varphi}.
\end{equation} 

In this way, the system of Eqs~(\ref{singlet_K0}) is reduced to independent systems of four equations defining the Fourier components $\psi_{3m}, \psi_{4m}, \psi_{7m}$, and $\psi_{8m}$ for each $m$. We do not write them explicitly, so as not to clutter the paper. 

Generally speaking, it is possible to transform the problem further by reducing it to an equation for a single function. This turns out to be useful for the further analysis. If we introduce the function $\Phi_m(r)=\psi_{3m}(r)+\psi_{8m}(r)$, one can exclude all functions $\psi_{3m},\psi_{4m},\psi_{7m},\psi_{8m}$  and obtain a single equation for $\Phi_m(r)$, 
\begin{multline}\label{Phi_m}
 \hat{k}^4_m\Phi_m+[2\lambda+a^2-g_2(\varepsilon,r)]\hat{k}^2_m\Phi_m-2g_1(\varepsilon,r)\frac{d}{dr}\left(\hat{k}^2_m\Phi_m\right)\\ -(2\lambda+a^2)g_1(\varepsilon,r)\frac{d}{dr}\Phi_m+[1-\widetilde{\varepsilon}(r)^2-\lambda g_2(\varepsilon,r)]\Phi_m=0,
\end{multline}
where the following designations are used 
\begin{equation}
 \hat{k}^2_m=-\frac{d^2}{dr^2}-\frac{1}{r}\frac{d}{dr}+\frac{m^2}{r^2},\quad \widetilde{\varepsilon}(r)=\dfrac{\varepsilon}{2}-v(r),
\end{equation}
\begin{equation}
 g_1(\varepsilon,r)=\frac{v'}{\widetilde{\varepsilon}},\quad g_2(\varepsilon,r)=\frac{2{v'}^2}{\widetilde{\varepsilon}^2}+\frac{v''}{\widetilde{\varepsilon}}+\frac{v'}{r\widetilde{\varepsilon}},
\end{equation} 
$v'$ and $v''$ are the potential derivatives.

The functions $\psi_{3m}$, $\psi_{4m}$, $\psi_{7m}$, and $\psi_{8m}$ are expressed through $\Phi_m(r)$ as follows:
\begin{align}
 \psi_{3m,8m}(r)&=\frac{1}{2}\left[1\mp\frac{1-\hat{k}_m^2}{\widetilde{\varepsilon}}\right]\Phi_m(r),\\ \psi_{4m,7m}(r)&=-\frac{ia}{2\widetilde{\varepsilon}}\left(\frac{d}{dr}\mp\frac{m}{r}\right)\Phi_m(r).
\end{align}

Equation~(\ref{Phi_m}) can be quite simply analyzed. It is seen that the equation has a singular point in which $\varepsilon-2v(r_c)=0$. If the interaction is repulsive, $v(r)>0$, the singularity exists only for $\varepsilon>0$. In this case, the solution can be analyzed by the expansion of the function $\Phi_m(r)$ near the singular point: $\Phi_m(r)=|r-r_c|^{\lambda}\sum_l a_l(r-r_c)^l$. In this way we come to the conclusion that $\Phi_m(r)$ and the components of the spinor $\Psi_{2m}(r)$ do not diverge at the point $r=r_c$. This fact allows one to further simplify the problem by using a model potential.

\subsection{Model steplike potential}
\label{model}

The physical understanding of the structure of two-particle states and their spectrum can be obtained by considering a model potential $v(r)$, which has the basic properties of the real potential of the pair interaction. As a model potential we choose a steplike function, 
\begin{equation}
 v(r)=\left\{ 
 \begin{array}{ll}
  v_0,& r<r_0,\\
  0,& r>r_0,
 \end{array}\right.
\end{equation} 
which is widely used and usually gives a good effective description of a more general class of short-range potentials.  

When using the steplike potential, an important point is to obtain matching conditions for the wave functions at the radius $r=r_0$. They should be obtained by integrating the full equations defining $\Psi_{j,\mathbf{K}}$ over the transition region, $|r-r_0|<\delta$, assuming that $v(r)$ is a finite value. Finally, the limit $\delta\to 0$ should be taken.

In this way, we arrive at the following matching equations in the case $\mathbf{K}=0$. For the singletlike states, one obtains
\begin{equation}
 \begin{array}{rl}\label{match_cond_singlet}
 \psi_{3m}\bigm|_-^+&=0,\\
 \psi_{8m}\bigm|_-^+&=0,\\
 2\frac{d\psi_{3m}}{dr}+ia(\psi_{4m}+\psi_{7m})\biggm|_-^+&=0,\\
 \frac{d\psi_{3m}}{dr}+\frac{d\psi_{8m}}{dr}\biggm|_-^+&=0.
\end{array}
\end{equation} 

It is interesting to note that the function $\Phi_m(r)$ is continuous at $r=r_0$. 

The same approach can be used for the tripletlike states. To be specific, we consider the state $\Psi_{1,\mathbf{K}}(\mathbf{r})$ at $\mathbf{K}=0$. The components $\psi_1(\mathbf{r})$, $\psi_2(\mathbf{r})$, $\psi_5(\mathbf{r})$, and $\psi_6(\mathbf{r})$ of the envelope function spinor are defined by the following equations:
\begin{equation}\label{triplet_K0}
\left\{
\begin{array}{rl}
 \left[2v(r)\!-\!\varepsilon\!-\!2\!+\!2\hat{k}^2\right]\psi_1(\mathbf{r})\!-\!a\hat{k}_+\psi_2(\mathbf{r})\!+\!a\hat{k}_+\psi_5(\mathbf{r}) &\!=0\\
 -a\hat{k}_-\psi_1(\mathbf{r})+\left[2v(r)-\varepsilon\right]\psi_2(\mathbf{r})+a\hat{k}_+\psi_6(\mathbf{r}) &\!=0\\
 a\hat{k}_-\psi_1(\mathbf{r})+\left[2v(r)-\varepsilon\right]\psi_5(\mathbf{r})-a\hat{k}_+\psi_6(\mathbf{r}) &\!=0\\
 a\hat{k}_-\psi_2(\mathbf{r})\!-\!a\hat{k}_-\psi_5(\mathbf{r})\!+\!\left[2v(r)\!-\!\varepsilon\!+\!2\!-\!2\hat{k}^2\right]\psi_6(\mathbf{r}) &\!=0\,.
\end{array}\right. 
\end{equation} 
It seen that the components $\psi_5(\mathbf{r})$ and $\psi_2(\mathbf{r})$ are connected by a simple relation: $\psi_2(\mathbf{r})=-\psi_5(\mathbf{r})$. 

In the polar coordinates, $\Psi_{1,K=0}$ is presented in the form of the Fourier series:
\begin{equation}\label{Psi1_F}
 \Psi_{1,K=0}(\mathbf{r})\!=\!\sum\limits_m \Psi_{1m} e^{im\varphi}\!=\!\sum\limits_m
 \begin{pmatrix}
 \psi_{1m}(r) e^{i\varphi} \\ \psi_{2m}(r) \\ \psi_{5m}(r) \\ \psi_{6m}(r) e^{-i\varphi}
 \end{pmatrix}
 e^{im\varphi}.
\end{equation} 

Equations defining the components $\psi_{1m}(r)$, $\psi_{2m}(r)$, $\psi_{5m}(r)$ and $\psi_{6m}(r)$ are easily obtained from Eq.~(\ref{triplet_K0}). The matching equations have the form:
\begin{equation}
 \begin{array}{rl}\label{match_cond_triplet}
 \psi_{1m}\bigm|_-^+&=0,\\
 \psi_{6m}\bigm|_-^+&=0,\\
 \frac{d\psi_{1m}}{dr}-ia\psi_{2m}\biggm|_-^+&=0,\\
 \frac{d\psi_{6m}}{dr}-ia\psi_{2m}\biggm|_-^+&=0.
\end{array}
\end{equation} 

Thus, Eqs.~(\ref{Psi2_F}), (\ref{singlet_K0}), and (\ref{match_cond_singlet}) fully define the singletlike state $\Psi_{2,K=0}(\mathbf{r})$. Correspondingly, Eqs.~(\ref{triplet_K0}), (\ref{Psi1_F}), and (\ref{match_cond_triplet}) define the tripletlike state $\Psi_{1,K=0}(\mathbf{r})$. Equations defining  $\Psi_{3,K=0}(\mathbf{r})$ and  $\Psi_{4,K=0}(\mathbf{r})$ can be obtained in a similar way. These equation are straightforwardly solved in Secs.~\ref{singlet}--\ref{trivial}, but before presenting the results of the calculations in detail, it is reasonable to stay on a qualitative picture of the bound state formation based on simplified models.
 
\section{A qualitative picture}
\label{qualitative}
In this section, we provide physical arguments, which qualitatively explain the mechanism of the bound state formation. They allow one also to better understand the main types of the bound states, which are obtained by solving the equations presented in the previous section. These arguments are derived from simplified models  with using different additional assumptions.

\subsection{Step potential}
First, consider the case of the interaction potential of the step form. Let us divide the space of the relative coordinate $\mathbf{r}$ into two regions: the interaction region, $r<r_0$, where $v(r)=v_0$, and the outer region, $r>r_0$, where the interaction is absent. 

In the outer region, the particles move freely, so that the spectrum of two particles with the zero total momentum contains two bands and zero-energy level:
\begin{align}
 \varepsilon=&\pm 2\sqrt{(\lambda+k^2)^2+a^2k^2}\,,\\
 \varepsilon=& 0\,.
\end{align}
The bands correspond to the particle configuration in which both particles have the energy in the conduction band or in the valence band. The zero-energy level is infinitely degenerate. This is a zero-energy mode, which corresponds to the case where the particles are in different bands with opposed momenta.

In the interaction region, the situation is very similar. Equation~(\ref{Schrodinger}) shows that the only effect of the inter-particle interaction on the two-particle spectrum is the shift of the energy by $2v_0$:
\begin{align}
 \varepsilon=&2v_0\pm 2\sqrt{(\lambda+k^2)^2+a^2k^2}\,,\\
 \varepsilon=& 2v_0\,.
\end{align}

\begin{figure}
\centerline{\includegraphics[width=.9\linewidth]{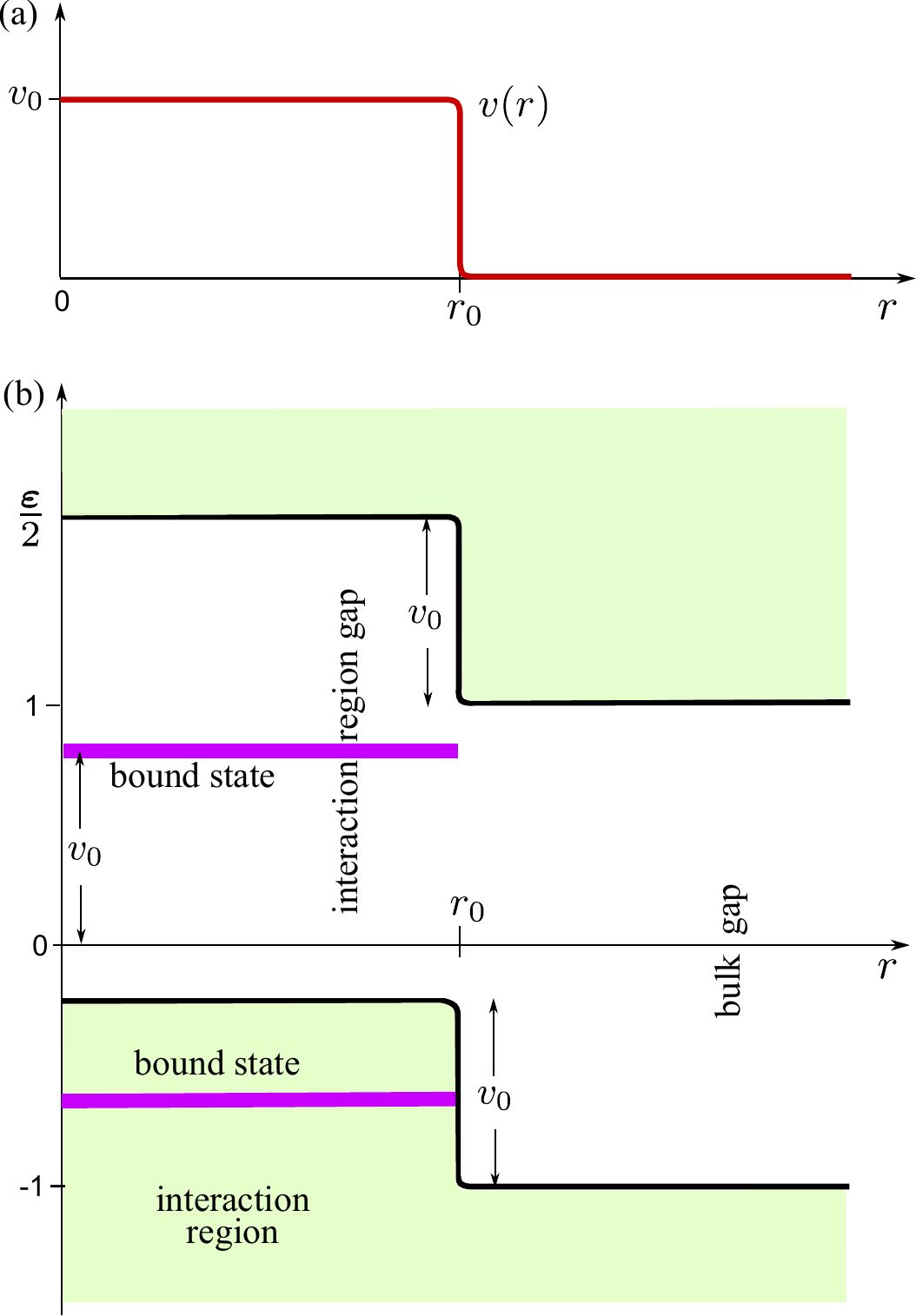}}
\caption{(Color online) The illustration of the mechanism of the bound state formation. (a) The pair interaction potential $v(r)$ versus the radius $r$. (b) Two-particle energy as a function of the radius. Propagating solutions exist in the light-green (darkened) areas. Uncolored areas are classically inaccessible. Wide violet lines indicate the energy at which bound states can be formed.}
\label{f_qualitative}
\end{figure}

Now imagine the energy diagram of the two-particle system in the space of the relative coordinate, see Fig.~\ref{f_qualitative}. Here, the areas in which there are propagating solutions of the Schr\"odinger equation, are colored in light green (darkened). Uncolored areas are classically inaccessible for the particles. Inside them, the solutions decay. It is obvious that in the energy interval $-2<\varepsilon<2(-1+v_0)$, the propagating solutions exist in the interaction region. In the outer region, the wave function decays if $\varepsilon<2$. The propagating solutions can interfere within the interaction region to form a two-particle bound state.

Another possibility to realize propagating solutions in the interaction region appears near the energy level of the zero-energy mode $\varepsilon\approx 2v_0$ in the interaction region, $r<r_0$. At this energy level, the wave functions decay outside the interaction region when $v_0<1$. In this way, localized two-particle states can also arise.

One can say that an effective quantum dot is formed in the interaction region where the bound states can be formed in the energy interval $-2<\varepsilon<2(-1+v_0)$ and near the energy $\varepsilon\approx 2v_0$.
 
Of course, these arguments are very qualitative, but in many respects they are true. The qualitative results agree with the rigorous solutions given in the following sections.

\subsection{A simplified model}
Another qualitative approach that illustrates the possibility of the bound state formation is based on the simplified model, which was used in early works on electron systems with inverted band structure~\cite{volkov1985pis,KorenmanDrewPRB1987}. In terms of the BHZ model, the simplified model neglects the term $Bk^2$ in the diagonal elements of the $\hat{h(k)}$ matrix in the Hamiltonian $\hat{H}_0$, see Eqs~(\ref{H_0}) and (\ref{h(k)}). This simplification is usually justified by the fact that the calculations are carried out within the $\mathbf{kp}$ theory where $k$ is supposed to be small. The system, reduced in such a way, in many cases leads to qualitatively correct results, but something is lost in it. The idea to qualitatively analyze localized states within two-band model by reducing it to an effectively single-band model goes back to the old work of L.V.~Keldysh~\cite{keldysh1964deep}, where this idea is applied to single-particle states localized at an impurity.

This simplification reduces the problem to the two-particle Dirac equation, which is well known in the quantum field theory. In recent years, the two-particle Dirac equation with a simplified electron-interaction interaction was adapted to the narrow-gap and gapless electronic systems in graphene and carbon nanotubes~\cite{SabioPRB2010,HartmannPRB2011,LeePRB2012,BermanPRA2013,HartmannAIP2016}.

In this way, we arrive at the following results. To be specific we consider only the singletlike states and turn to Eq.~(\ref{Phi_m}) for the function $\Phi_m(r)$. Turning to the reduced model, we have to put $B=0$ in the BHZ Hamiltonian. Since $B$ is used in defining the dimensionless variables, introduced in Eq.~(\ref{dim_less_param}), one needs to go back to the dimensional quantities, which will be used only in Eqs.~(\ref{Phi_short})--(\ref{mid-gap_solution}) below. Finally, in the reduced model we arrive at the following equation instead of Eq.~(\ref{Phi_m}):
\begin{multline}\label{Phi_short}
 A^2\hat{k}^2\Phi_m-A^2\frac{2V'(r)}{E-2V(r)}\frac{d\Phi_m}{dr}+[EV(r)-V^2(r)]\Phi_m\\+\left(M^2-\frac{E^2}{4}\right)\Phi_m=0.
\end{multline} 
Note that a similar equation was used for the two-particle systems in Refs.~\cite{SabioPRB2010,HartmannPRB2011,LeePRB2012,BermanPRA2013,HartmannAIP2016}.

One can consider this equation as a single-particle problem, where the first term is the kinetic energy and the third term plays a role of an effective potential. Its sign depends on the energy and the magnitude of $V(r)$. So, the effective potential is negative  (i.e. attractive) when $E<0$. It can be negative also at $E>0$, if the real potential $V$ is high enough. Since the effective potential is attractive in a wide range of $E$, one can expect that bound states can be formed, at least in the case where the second term in Eq.~(\ref{Phi_short}) is not large [for example, when $V(r)$ is a slow varying function]. 

It is interesting to consider the same equation from other point of view. We redefine the potential so that it becomes positive (for example, by multiplying the equation by a number). In this case, the first term, which plays the role of kinetic energy, becomes negative and therefore the effective reduced mass of the two particles is negative.

This is well illustrated by considering Eq.~(\ref{Phi_short}) in the limiting case where $V(r)\ll |M|$, $E=-2|M|+2\Delta E$ and $\Delta E\ll |M|$.  Equation~(\ref{Phi_short}) takes the form
\begin{equation}
 -\frac{A^2}{2|M|}\hat{k}^2\Phi_m-\frac{A^2}{2M^2}V'(r)\Phi_m+V(r)\Phi_m=\Delta E\,\Phi_m\,.
\end{equation} 
It is seen that the effective reduced mass is negative and can be defined as $m^*=-\hbar^2|M|/A^2$. Thus, a bound state is formed by the positive potential. This is obvious, at least, if one neglects the second term, which really can be dropped if the characteristic length $l$ of the potential change is large, $l^2\gg A^2/(2|M|)$.

Another possibility for a bound state to appear arises because of the singularity of the second term in Eq.~(\ref{Phi_short}) in the point $r=r_0$ where $E=2V(r_0)$. Again, consider a simplified case where $E/2, V(r)\ll |M|$ and $E/2$ is smaller than the maximum value of the interaction potential. If $V(r)$ is a monotonic function, there is one singular point. We focus on the solution of Eq.~(\ref{Phi_short}) near the point $r=r_0$ by expanding the potential: $V(r)=E/2+V'(r_0)(r-r_0)+\dots$. In this case, Eq~(\ref{Phi_short}) takes the form
\begin{equation}
 \hat{k}^2\Phi_m+\frac{1}{r-r_0}\frac{d\Phi_m}{dr}+\frac{M^2}{A^2}\Phi_m=0\,.
\end{equation} 
This equation is easily solved in the vicinity of the point $r=r_0$. For $m=0$, one obtains the following solution
\begin{equation}\label{mid-gap_solution}
 \Phi(r)\simeq \mathrm{const}\,|r-r_0|\,K_1\left(\left|\frac{M}{A}(r-r_0)\right|\right)\,,
\end{equation} 
where $K_1(z)$ is the modified Bessel function of the second kind.

This fact argues that there can be a solution localized near the point $r=r_0$, but it fails to determine the eigenenergy $E$ since the complete solution satisfying boundary conditions is not found. Nevertheless, the qualitative behavior of the wave function agrees with the results of Refs.~\cite{SabioPRB2010,LeePRB2012} where such a singularity was studied in the case of massless Dirac fermions and it was found that a quasibound state appears with the wave function effectively localized near $r=r_0$. The estimate~(\ref{mid-gap_solution}) qualitatively agrees also with the total solution of the problem, which will be presented in the next sections.

\section{Singletlike bound states}
\label{singlet}
In this section, the singletlike states are studied by the direct solution of Eq.~(\ref{singlet_gen}) in the case where the potential has the step form and $\mathbf{K}=0$ . First, we consider of the states $\Psi_{2,\mathbf{K=0}}(\mathbf{r})$. We find their spectrum and the spatial distribution of all components of the envelope function spinor. Then the results are generalized to the states $\Psi_{3,\mathbf{K=0}}(\mathbf{r})$, and finally we obtain the two-particle wave function which is antisymmetric with respect to the permutation of the particles. To be specific we consider below the case of the topological insulator ($\lambda=-1$). The topologically trivial case will be presented in Sec.~\ref{trivial}.

\subsection{Spectrum}
The states described by the wave function $\Psi_{2,\mathbf{K}=0}(\mathbf{r})$ (the index $\mathbf{K}=0$ is dropped hereinafter) are determined by Eqs.~(\ref{singlet_K0}). We solve these equations in the regions $r<r_0$ and $r>r_0$, and match the found functions at $r=r_0$ with using Eqs.~(\ref{match_cond_singlet}).

In the case of the step potential, Eqs.~(\ref{singlet_K0}) are easily solved in terms of the Bessel functions. The fundamental set of solutions for the components of the spinor $\Psi_2$ has the form:
\begin{equation}\label{Bessel_base}
 \begin{array}{ll}
  \psi_{3m}(r)=A^{\pm}_m \mathcal{F}_m(Q_{\pm}r), &\psi_{4m}(r)=B^{\pm}_m \mathcal{F}_{m+1}(Q_{\pm}r),\\ \psi_{7m}(r)=C^{\pm}_m \mathcal{F}_{m-1}(Q_{\pm}jr), &\psi_{8m}(r)=D^{\pm}_m \mathcal{F}_m(Q_{\pm}r),
 \end{array}
\end{equation} 
where the wave numbers $Q_{\pm}$ are the roots of the dispersion equation, which has a unified form in both regions:
\begin{equation}
 \widetilde{\varepsilon}^{\,2}\left[\widetilde{\varepsilon}^{\,2}-(1-Q^2)^2-a^2Q^2\right]=0,
\end{equation} 
where $\widetilde{\varepsilon}$ takes different values for the interaction region and the outer region,
\begin{equation}
 \widetilde{\varepsilon}=\left\{
 \begin{array}{ll}
  \varepsilon_0 - v_0\,,\; & r<r_0,\\
  \varepsilon_0\,,\; & r>r_0.
 \end{array}
 \right.
\end{equation} 
For convenience we have denoted here $\varepsilon_0\equiv \varepsilon/2$, which is the energy of an electron pair per particle.

\begin{figure}
\centerline{\includegraphics[width=.9\linewidth]{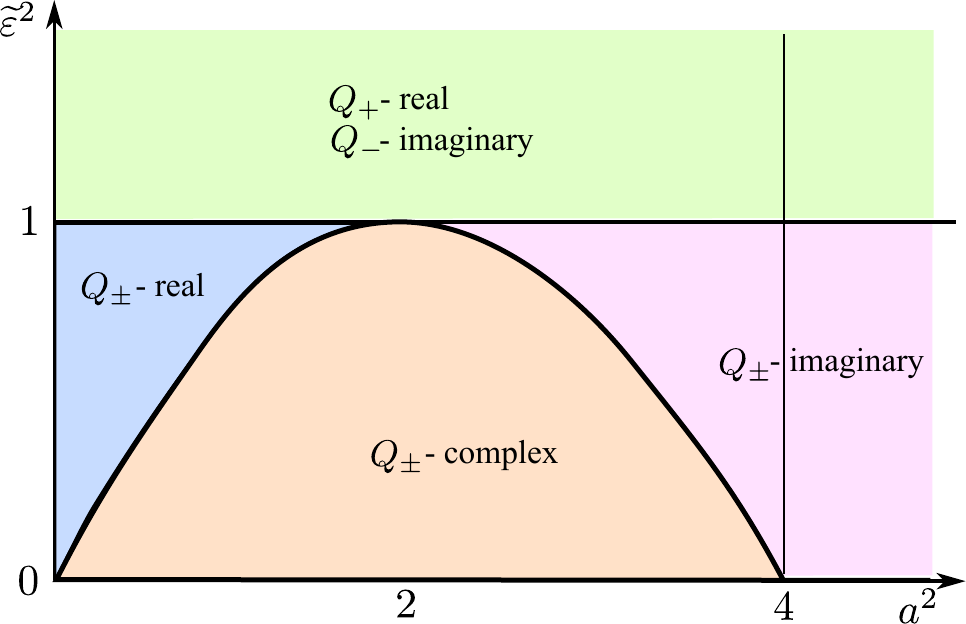}}
\caption{(Color online) Map of possible values of $Q_{\pm}$ on the plane $(a^2,\widetilde{\varepsilon})$.}
\label{f-Q}
\end{figure}

The explicit expression for $Q_{\pm}$ reads
\begin{equation}\label{Q}
 Q_{\pm}=\sqrt{1-\frac{a^2}{2}\pm\sqrt{a^2\left(\frac{a^2}{4}-1\right)+\widetilde{\varepsilon}^2}}.
\end{equation} 
In what follows it is important that $Q_{\pm}$ can be real, imaginary or complex, depending on the parameter $a$ and the energy $\widetilde{\varepsilon}$. The map of possible values of $Q_{\pm}$ on the plane $(a^2,\widetilde{\varepsilon})$ is shown in Fig.~\ref{f-Q}.

In Eq.~(\ref{Bessel_base}), $\mathcal{F}_m(Q_{\pm}r)$ is a fundamental solution of the Bessel equation. $\mathcal{F}_m(Q_{\pm}r)$ can be written as any pair of the Bessel functions: $J_m(Q_{\pm}r)$ and $Y_m(Q_{\pm}r)$; $I_m(Q_{\pm}r)$ and $K_m(Q_{\pm}r)$; $H_m^{(1),(2)}(Q_{\pm}r)$; etc.. The choice of the pair of Bessel functions in a specific case is determined by the values of $Q_{\pm}$ at given $a$ and $\widetilde{\varepsilon}$ (in accordance with the map in Fig.~\ref{f-Q}), and by the behavior of the Bessel function at $r\to 0$ and $r\to \infty$.

To be specific consider the case where $a^2>4$. 

In the energy interval $-1<\varepsilon_0<-1+v_0$, the solution of Eqs~(\ref{singlet_K0}) can be presented in the following form:\\

(i) at $r<r_0$,
\begin{equation}\label{wave_func-}
 \begin{array}{ll}
 \psi_{3m}=& A_+J_m(k_+r)+A_-I_m(k_-r),\\
 \psi_{4m}=& A_+\mathcal{B}_+J_{m+1}(k_+r)+A_-\mathcal{B}_-I_{m+1}(k_-r),\\
 \psi_{7m}=& A_+\mathcal{C}_+J_{m-1}(k_+r)+A_-\mathcal{C}_-I_{m-1}(k_-r),\\
 \psi_{8m}=& A_+\mathcal{D}_+J_m(k_+r)+A_-\mathcal{D}_-I_m(k_-r),
 \end{array}
\end{equation} 
where
\begin{equation}
 \begin{array}{ll}
 \mathcal{B}_{\pm}=& i\dfrac{\varepsilon_0-v_0+1\mp k_\pm^2}{ak_\pm},\\
 \mathcal{C}_{\pm}=& -i\dfrac{\varepsilon_0-v_0+1\mp k_\pm^2}{ak_\pm},\\
 \mathcal{D}_{\pm}=& \dfrac{\varepsilon_0-v_0+1\mp k_\pm^2}{\varepsilon_0-v_0-1\pm k_\pm^2},
 \end{array}
\end{equation} 
and 
\begin{equation}\label{k_pm}
 k_{\pm}=\sqrt{\pm\!\left(1\!-\!\frac{a^2}{2}\right)\!+\!\sqrt{a^2\left(\frac{a^2}{4}\!-\!1\right)+(\varepsilon_0\!-\!v_0)^2}};
\end{equation} 

(ii) at $r>r_0$,
\begin{equation}\label{wave_func+}
 \begin{array}{ll}
 \psi_{3m}=& B_+K_m(\kappa_+r)+B_-K_m(\kappa_-r),\\
 \psi_{4m}=& B_+\mathcal{K}_+K_{m+1}(\kappa_+r)+B_-\mathcal{K}_-K_{m+1}(\kappa_-r),\\
 \psi_{7m}=& B_+\mathcal{L}_+K_{m-1}(\kappa_+r)+B_-\mathcal{L}_-K_{m-1}(\kappa_-r),\\
 \psi_{8m}=& B_+\mathcal{M}_+K_m(\kappa_+r)+B_-\mathcal{M}_-K_m(\kappa_-r),
 \end{array}
\end{equation} 
where
\begin{equation}
 \begin{array}{rl}
 \mathcal{K}_{\pm}=\mathcal{L}_{\pm}&=-i\dfrac{\varepsilon_0+1+\kappa_\pm^2}{a\kappa_\pm},\\
 \mathcal{M}_{\pm}=& \dfrac{\varepsilon_0+1+\kappa_\pm^2}{\varepsilon_0-1-\kappa_\pm^2},
 \end{array}
\end{equation}
and
\begin{equation}
 \kappa_{\pm}=\sqrt{-1+\frac{a^2}{2}\pm \sqrt{a^2\left(\frac{a^2}{4}-1\right)+\varepsilon_0^2}}.
\end{equation} 

Now the functions~(\ref{wave_func-}) and (\ref{wave_func+}) should be matched at the boundary $r=r_0$. Using Eqs.~(\ref{match_cond_singlet}), we get a homogeneous equation system for the coefficients $A_+, A_-, B_+, B_-$. The equations are very cumbersome, so we do not give them and subsequent equations in an explicit form. The determinant $\mathfrak{D}$ of this equation system is a function of the energy $\varepsilon_0$ and the parameters $a, v_0, r_0, m$. The eigenenergies are determined by the equation 
\begin{equation}\label{DD}
\mathfrak{D}(\varepsilon_0;a,v_0,r_0,m)=0.                                          
\end{equation}  
It turns out that this equation has several solutions:
\begin{equation}
\varepsilon_{n,m}^{(s)}=2\varepsilon_{0,m}(a,v_0,r_0),                                        
\end{equation}
where $n$ is a root number at given parameters $a$, $v_0$, $r_0$, and the angular number $m$. One can say that $n$ is the radial quantum number. The upper symbol indicates that this is a singletlike state. 

In the energy interval $-1+v_0<\varepsilon_0<1$, the solution of Eqs~(\ref{singlet_K0}) differs from that considered above since in the interaction region both roots $Q_{\pm}$, see Eq.~(\ref{Q}), are imaginary. Therefore the solution in the region $r<r_0$ should be composed of the Bessel functions $I_m(|Q_{\pm}|)$. This is the only difference from Eqs~(\ref{wave_func-}) and (\ref{wave_func+}). Moreover, it is clear that Eqs~(\ref{wave_func-}) and (\ref{wave_func+}) are formally correct in the interval $-1+v_0<\varepsilon_0<1$, if one considers $k_+$ as a complex number.

Equation~(\ref{DD}) for the eigenvalues of the energy is very cumbersome and complicated. In this paper, we solve it numerically. This approach enables us to find solutions for a finite value of fundamentally important parameters $v_0$ and $r_0$. As a result we demonstrate the presence of bound states and the main features of their spectrum. The main result of these studies is that two-particle bound states exist in a wide range of the parameters $v_0$, $r_0$ and $a$. Energy levels of the bound states lie in the gap of the band spectrum. 

The analysis has shown that the spectrum of the two-particle states is more complicate than one could expect from the qualitative arguments of Sec.~\ref{qualitative}. In this section we have restrict ourselves to the bound states with zero angular number and the region of the parameter $|a|>2$. In this case, the calculations turn out to be more simple and the results seem to be quite general. Qualitatively new features of the bound states are expected when the parameter $|a|<2$. This case will be studied in Sec.~\ref{invert}. 

\begin{figure}
\centerline{\includegraphics[width=.9\linewidth]{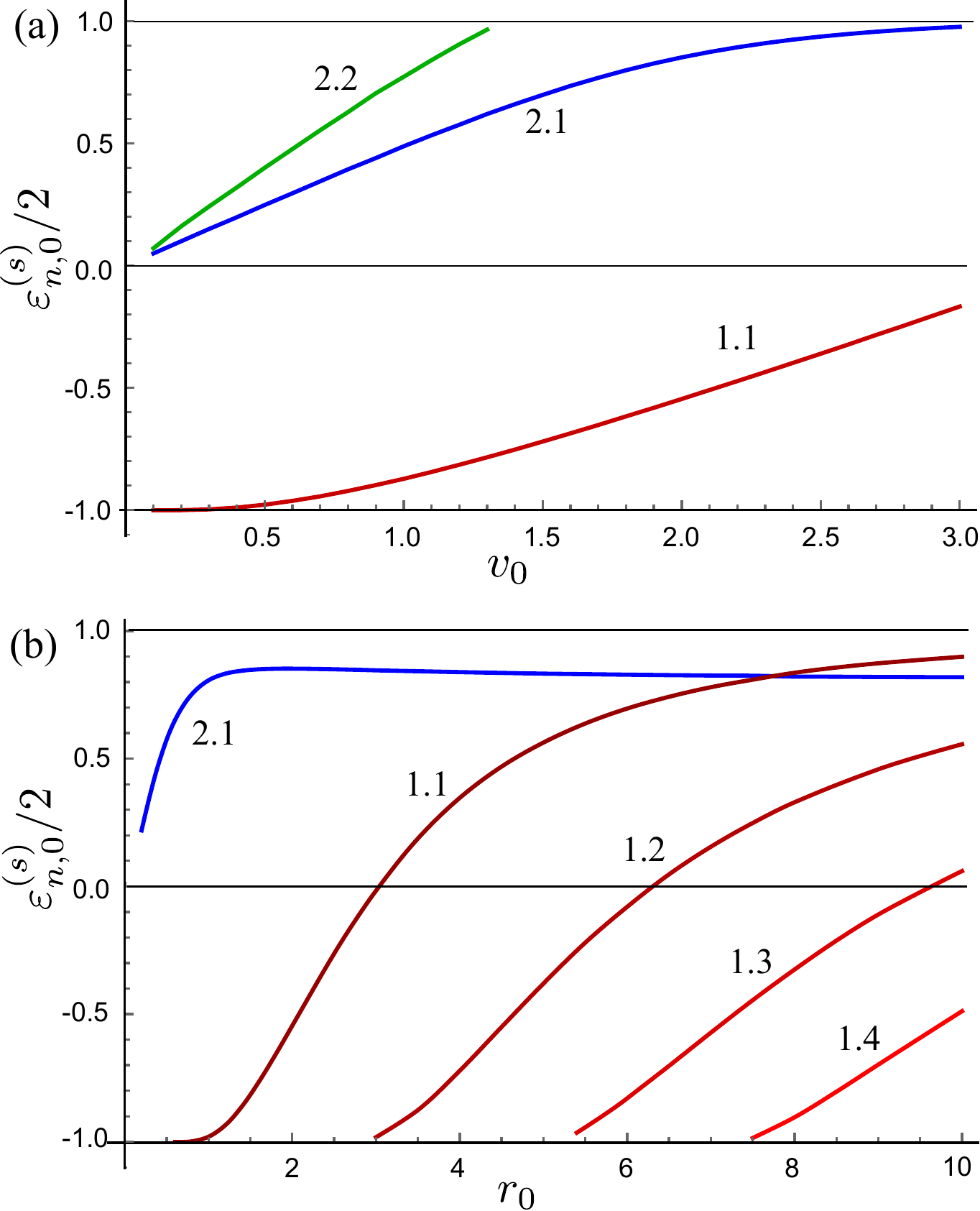}}
\caption{(Color online) The spectrum of the singletlike bound states. (a) The bound-state energy $\varepsilon$ as a function of the interaction potential $v_0$. The parameters used in the calculations: $a=2.1$, $r_0=2.0$, $m=0$. (b) The bound-state energy $\varepsilon$ as a function of the interaction radius $r_0$. Lines 1.1, 1.2, 1.3, 1.4 refer to the states of the first group. Line 2.1 refers to the states of the second group. The parameters used in the calculations: $a=2.1$, $v_0=2.0$, $m=0$.}
\label{f-singlet_spectrum}
\end{figure}

There are two groups of the bound states in accordance with the qualitative arguments of Sec.~\ref{qualitative}. They are classified by the energy that a bound state has at low interaction potential. The energy levels of the first group appear at the bottom of the gap of the two-particle band spectrum and then rise with increasing $v_0$. The bound states of the second group have the energy near the center of the band gap at low $v_0$. The behavior of the energy levels of the first and second groups with the increase of the interaction potential is illustrated in Fig.~\ref{f-singlet_spectrum}(a) in the case where the interaction radius $r_0$ is not large, so that only one state of the first group exists, when $m=0$. In contract, the second group contains two states with $m=0$. Of course, the states with $m\ne 0$ also exist in both groups, but the dependence of their energy on the parameters $v_0$ and $r_0$ is more complicated than one might expect at first glance.

It is obvious that this classification is justified only at low interaction potential. When $v$ is comparable with the band gap, this classification is very conventional and little constructive. Nevertheless, we will stick to it to trace the evolution of the bound states with increasing the interaction potential.

The dependence of the bound state energy on the interaction radius $r_0$ is illustrated in Fig.~\ref{f-singlet_spectrum}(b). The states of the first group behave as follows. With increasing $r_0$, new roots of the determinant $\mathfrak{D}(\varepsilon;a,v_0,r_0,m)$ successively appear at the bottom of the gap. This remembers the usual picture of quantization in a quantum dot. In the case we are studying, such a quantum dot is effectively formed by the interaction potential as illustrated in Fig.~\ref{f_qualitative}. Of course, the quantization conditions are very different from those in ordinary quantum dots in one-band model with a quadratic dispersion. 

The states of the second group show a completely different behavior. With the increase of $r_0$, no additional roots appear with zero angular number. This feature could be understood as a result of the fact that the quantum state is localized along the perimeter of the effective quantum dot similarly to an edge state, rather than inside it. In this case, only the angular motion is quantized.

Thus, in order to elucidate the mechanism of the bound state formation it is interesting to analyze the spatial distribution of the electron density and the density of all components of the envelope function spinor. 

\subsection{Electronic structure of the bound states}
The envelope functions $\psi_3(r)$, $\psi_4(r)$, $\psi_7(r)$ and $\psi_8(r)$ in the state $\Psi_2(\mathbf{r})$ can be calculated straightforwardly with using Eqs~(\ref{wave_func-}), (\ref{wave_func+}) and coefficients $A_+$, $A_-$, $B_+$, and $B_-$. 

We begin with the states of the first group. 

\subsubsection{First group of bound states}
\begin{figure}
\centerline{\includegraphics[width=1.\linewidth]{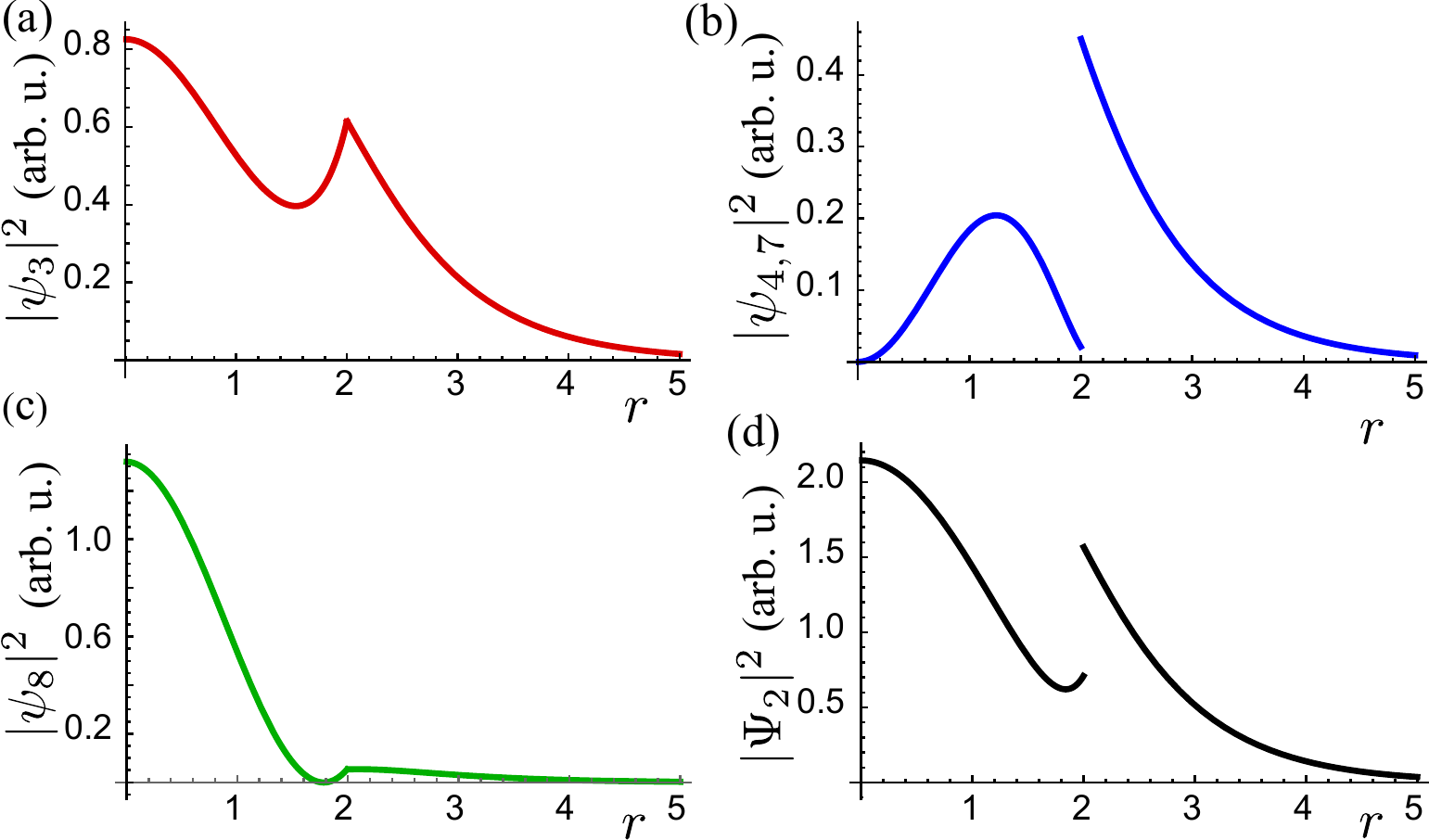}}
\caption{(Color online) The radial distribution of the spinor component densities in the singletlike state, $\Psi_2(\mathbf{r})$, of the first group: (a) the component $\psi_3(r)$,  (b) the components $\psi_4(r)$ and $\psi_7(r)$, (c) the component $\psi_8(r)$, (d) the total density $|\Psi_2(r)|^2$. The parameters used in the calculations: $a=2.1$, $v_0=2.0$, $r_0=2.0$, $m=0$, $\varepsilon/2= -0.545259031$.}
\label{f-singlet1_w-func}
\end{figure}

The radial distribution of the density of all spinor components [$\psi_3(r)$, $\psi_4(r)$, $\psi_7(r)$, and $\psi_8(r)$] is shown in Figs.~\ref{f-singlet1_w-func}(a)--(c) for the bound state corresponding to the line 1.1 in Fig.~\ref{f-singlet_spectrum}(a). Here it should be noted that in the case of zero angular number, $m=0$, the spinor components $\psi_4(r)$ and $\psi_7(r)$ coincide though, in general, $\psi_4(r)\ne \psi_7(r)$. The discontinuity of some envelope functions or their derivatives at $r=r_0$ originates from the singularity of the potential. A separate investigation of the solutions in the vicinity of the point $r=r_0$ in the case where the potential is approximated by a linear function with large gradient, shows that the wave function is also continuous but sharply changes.

It is seen that the spinor components $\psi_3(r)$ and $\psi_8(r)$ have the largest amplitude. They describe the contribution of the two-particle basis states $|E\uparrow E\downarrow\rangle$ and $|H\uparrow H\downarrow\rangle$ into the total wave function, respectively. Thus, the states of this group are formed mainly by those orbital components, in which both particles are in the electron band or in the hole band. The contribution of the mixed components $|E\uparrow H\downarrow\rangle$ and $|H\uparrow E\downarrow\rangle$ is small. In addition, the amplitude of the mixed components strongly decreases with decreasing $v_0$. Another conclusion is that the electron density is distributed in the volume of the effective quantum dot, though there is also a small density located at the edge. 

The radial distribution of the total density $|\Psi_2(r)|^2=|\psi_3(r)|^2+|\psi_4(r)|^2+|\psi_7(r)|^2+|\psi_8(r)|^2$ is shown in Fig.~\ref{f-singlet1_w-func}(d).

Now we turn to the second group of the bound states.

\subsubsection{Second group of bound states}
\begin{figure}
\centerline{\includegraphics[width=1.\linewidth]{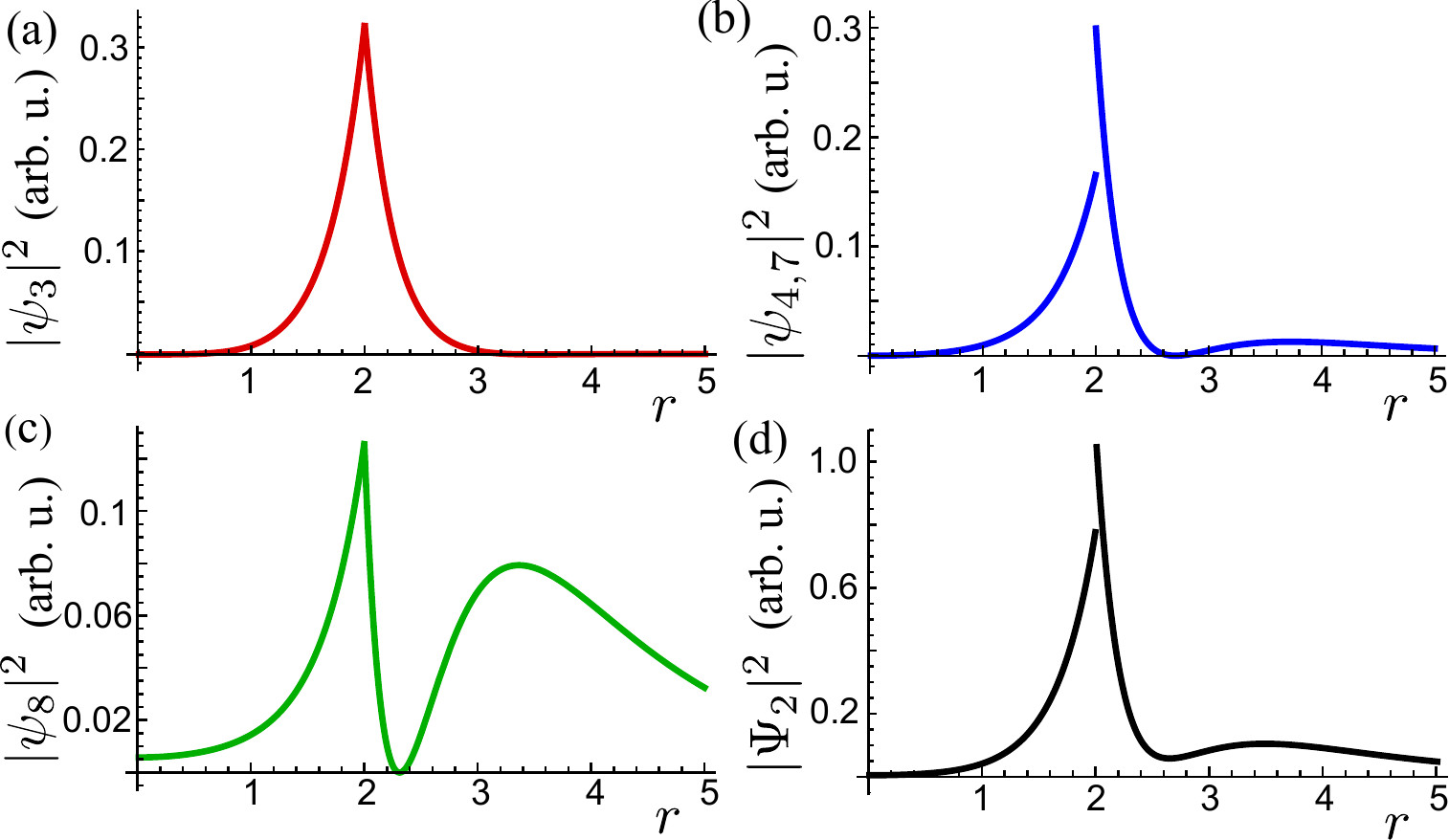}}
\caption{(Color online) The radial distribution of the spinor components in the singletlike state, $\Psi_2(\mathbf{r})$, of the second group: (a) the component $\psi_3(r)$,  (b) the components $\psi_4(r)$ and $\psi_7(r)$, (c) the component $\psi_8(r)$, (d) the total density $|\Psi_2(r)|^2$. The parameters used in the calculations: $a=2.1$, $v_0=2.0$, $r_0=2.0$, $m=0$, $\varepsilon/2= 0.8515680419$.}
\label{f-singlet2_w-func}
\end{figure}

The radial distribution of the densities of the spinor components is shown in Fig.~\ref{f-singlet2_w-func}(a,b,c) for the bound state shown by the line 2.1 in Fig.~\ref{f-singlet_spectrum}. 

In this group of states, the amplitude of the components $\psi_4(r)$ and $\psi_7(r)$ noticeably increases as compared with the first group states. These components represent the contribution of the mixed states of the electron and hole bands ($|E\uparrow H\downarrow\rangle$ and $|H\uparrow E\downarrow\rangle$) to the total wave function. However, the main distinction from the first group states is that the particle density is concentrated at the edge of the effective quantum dot. Hence, this state can be considered as a kind of edge states.

\subsection{Two-electron wave function}
The true wave function of the two electrons is to be antisymmetric with respect to the permutation of particles. The wave function described by the spinor $\Psi_{2,\mathbf{K}}(\mathbf{r})$ does not satisfy this requirement. Therefore the two-electron wave function should be presented in the form:
\begin{equation}
 \Psi^{(s)}(1,2)=\frac{1}{\sqrt{2}}\left[\Psi_2(1,2)-\Psi_2(2,1)\right],
\end{equation} 
where the arguments $1,2$ denote the coordinates of two electrons and $\Psi_2(1,2)$ is the wave function of the state described by the envelope function spinor $\Psi_2(\mathbf{r})$. The permutation of the particles includes both the replacement $\mathbf{r}\to -\mathbf{r}$ and the interchange of the particle coordinates in the two-particle basis functions, Eq.~(\ref{basis_f}). Taking into account this fact, we arrive at the following wave function:
\begin{equation}\label{true_singlet_WF}
\begin{split}
 \Psi_m^{(s)}(1,2)&= C\left[\psi_{3m}(r)\left(|E\uparrow E\downarrow\rangle-|E\downarrow E\uparrow\rangle\right)\right.\\
 &\left.+ \psi_{4m}(r)e^{i\varphi}\left(|E\uparrow H\downarrow\rangle+|H\downarrow E\uparrow\rangle\right)\right.\\
 &\left.+ \psi_{7m}(r)e^{-i\varphi}\left(|H\uparrow E\downarrow\rangle+|E\downarrow H\uparrow\rangle\right)\right.\\
 &\left.+ \psi_{8m}(r)\left(|H\uparrow H\downarrow\rangle-|H\downarrow H\uparrow\rangle\right)\right].
\end{split}
\end{equation}

As it is seen, the wave function~(\ref{true_singlet_WF}) can not be factorized into orbital and spin functions. Therefore, this state can not be called a singlet state in the usual sense. Nevertheless we continue to use this nonstrict term.

We finish this section by considering another singletlike state $\Psi_3(\mathbf{r})$. Straightforward calculations show that the components of this spinor [$\psi_9(\mathbf{r})$,  $\psi_{10}(\mathbf{r})$, $\psi_{13}(\mathbf{r})$, and $\psi_{14}(\mathbf{r})$] are determined by the same equations as the components of $\Psi_2(\mathbf{r})$. One can show that the components of $\Psi_3(\mathbf{r})$ are connected with those of $\Psi_2(\mathbf{r})$ by the following replacement: $\psi_9\to \psi_3$, $\psi_{10}\to -\psi_7$, $\psi_{13}\to -\psi_4$ and $\psi_{14}\to \psi_8$. Taking into account this replacement together with the replacements in the basis functions, it is easy to see that the spinor wave function $\Psi_3(\mathbf{r})$ differs from the wave function $\Psi_2(\mathbf{r})$ simply by the permutation of the particles. Thus, the quantum state described by the spinor $\Psi_3(\mathbf{r})$ coincides with the already-studied state, $\Psi_m^{(s)}(1,2)$.

\section{Tripletlike bound states}
\label{triplet}
The tripletlike states are studied similarly to the singletlike ones with using the same simplifications. Therefore, we do not go into the details and only present main results.

There are two tripletlike states: $\Psi_{1,\mathbf{K}}(\mathbf{r})$ and $\Psi_{4,\mathbf{K}}(\mathbf{r})$ which differ only in the direction of the spins. Since the system studied here has $S_z$ symmetry, the other properties of these states are the same. Below we consider only the state $\Psi_{1,\mathbf{K}}(\mathbf{r})$ at $\mathbf{K}=0$ and focus on the topologically nontrivial case as in the previous section.

\subsubsection{The spectrum}

\begin{figure}
\centerline{\includegraphics[width=.9\linewidth]{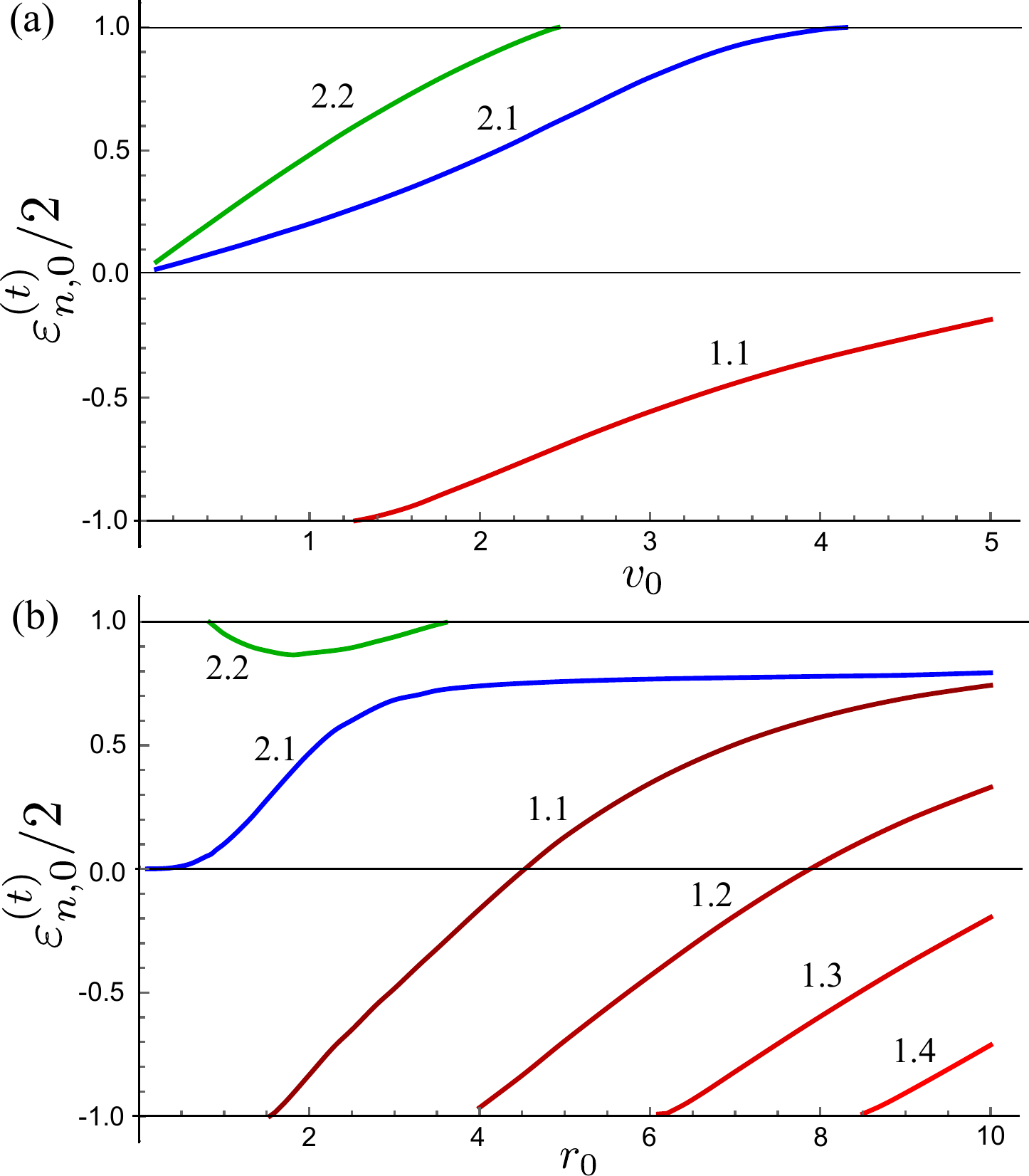}}
\caption{(Color online) The spectrum of the tripletlike bound states. (a) The  bound-state energy $\varepsilon$ as a function of the interaction potential $v_0$ at $r_0=2.0$. (b) The bound-state energy as a function of the interaction radius $r_0$ at $v_0=2.0$. Lines 1.1, 1.2, 1.3, 1.4 refer to the states of the first group. Lines 2.1 and 2.2 refer to the states of the second group. The parameters used in the calculations: $a=2.1$, $m=0$.}
\label{f-triplet_spectrum}
\end{figure}

The energy spectrum of the tripletlike bound states is generally similar to the spectrum of the singletlike states, but there are some differences in details. The tripletlike bound states can also be divided into two groups which differ in the energy at the low interaction potential. The states of the first group have the energy at the bottom of the band gap, while the energy of the second-group states lies near the center of the gap. The bound-state energy is determined by three parameters of the model ($a$, $v_0$ and $r_0$) and two quantum numbers: the radial quantum number $n$ and the angular quantum number $m$. The dependence of the energy on the potential amplitude is illustrated in Fig.~\ref{f-triplet_spectrum}(a) for $m=0$. The interaction radius is chosen so small that there is only one energy level of the states of the first group. In contrast, the second group contains two states even if $v_0$ is small. The evolution of the spectrum with increasing $r_0$ is shown in Fig.~\ref{f-triplet_spectrum}(b). It is seen that new states with $m=0$ arise only in the first group. They are characterized by the radial quantum number.

\subsubsection{Electronic structure of the bound states}
Electronic structure of the tripletlike bound states in many respects is also similar to that of the singletlike state, but there are many significant differences in the spatial distribution of the densities of the spinor components related to the electron and hole bands.

\begin{figure}
\centerline{\includegraphics[width=1.\linewidth]{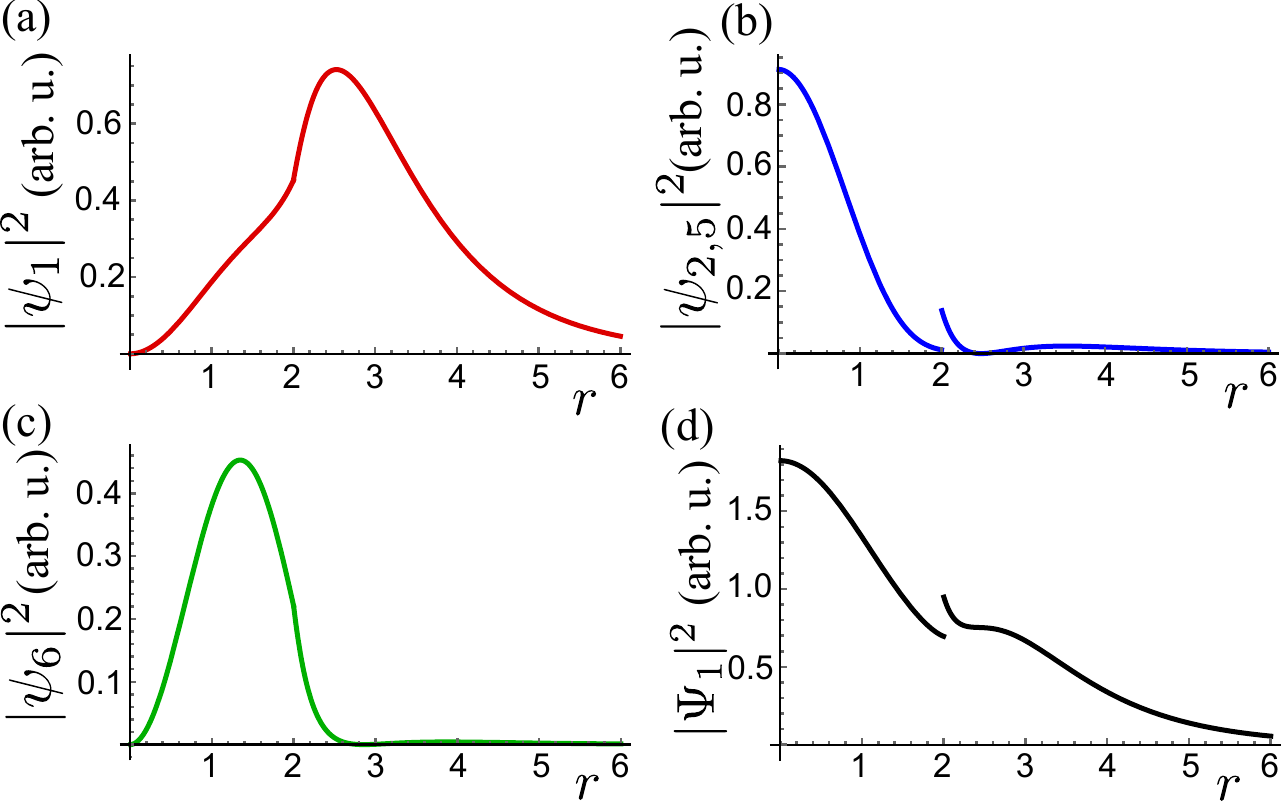}}
\caption{(Color online) The radial distribution of the spinor component densities in the tripletlike state of the first group: (a) the component $\psi_1(r)$,  (b) the components $\psi_2(r)$ and $\psi_5(r)$, (c) the component $\psi_6(r)$, (d) the total density $|\Psi_1(r)|^2$. The parameters used in the calculations: $a=2.1$, $v_0=2.0$, $r_0=2.0$, $m=0$, $\varepsilon/2= -0.831800948$.}
\label{f-triplet1_w-func}
\end{figure}

\begin{figure}
\centerline{\includegraphics[width=1.\linewidth]{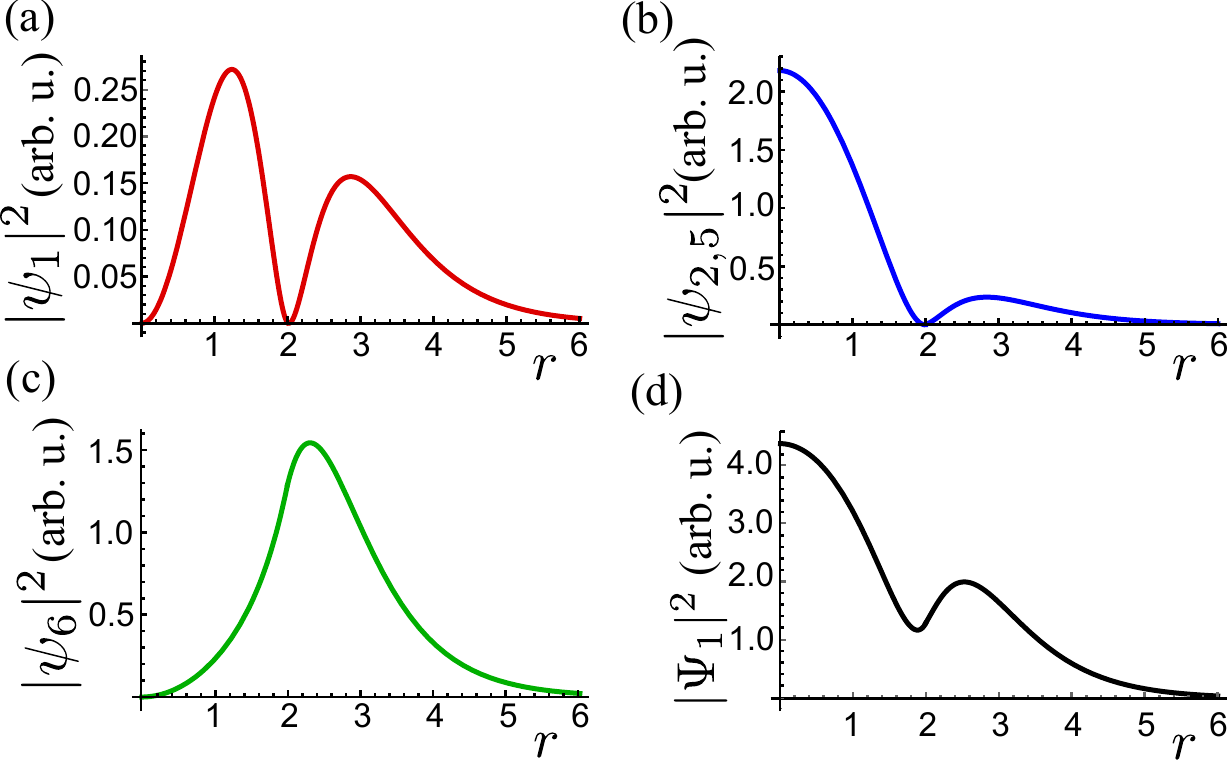}}
\caption{(Color online) The radial distribution of the spinor component densities in the tripletlike states of the second group. Panels (a)--(d) represent the spinor components in the state shown by the line 2.1 in Fig.~\ref{f-triplet_spectrum} at $\varepsilon/2= 0.4680414375$. Other parameters: $a=2.1$, $v_0=2.0$, $r_0=2.0$, $m=0$.}
\label{f-triplet21_w-func}
\end{figure}

\begin{figure}
\centerline{\includegraphics[width=1.\linewidth]{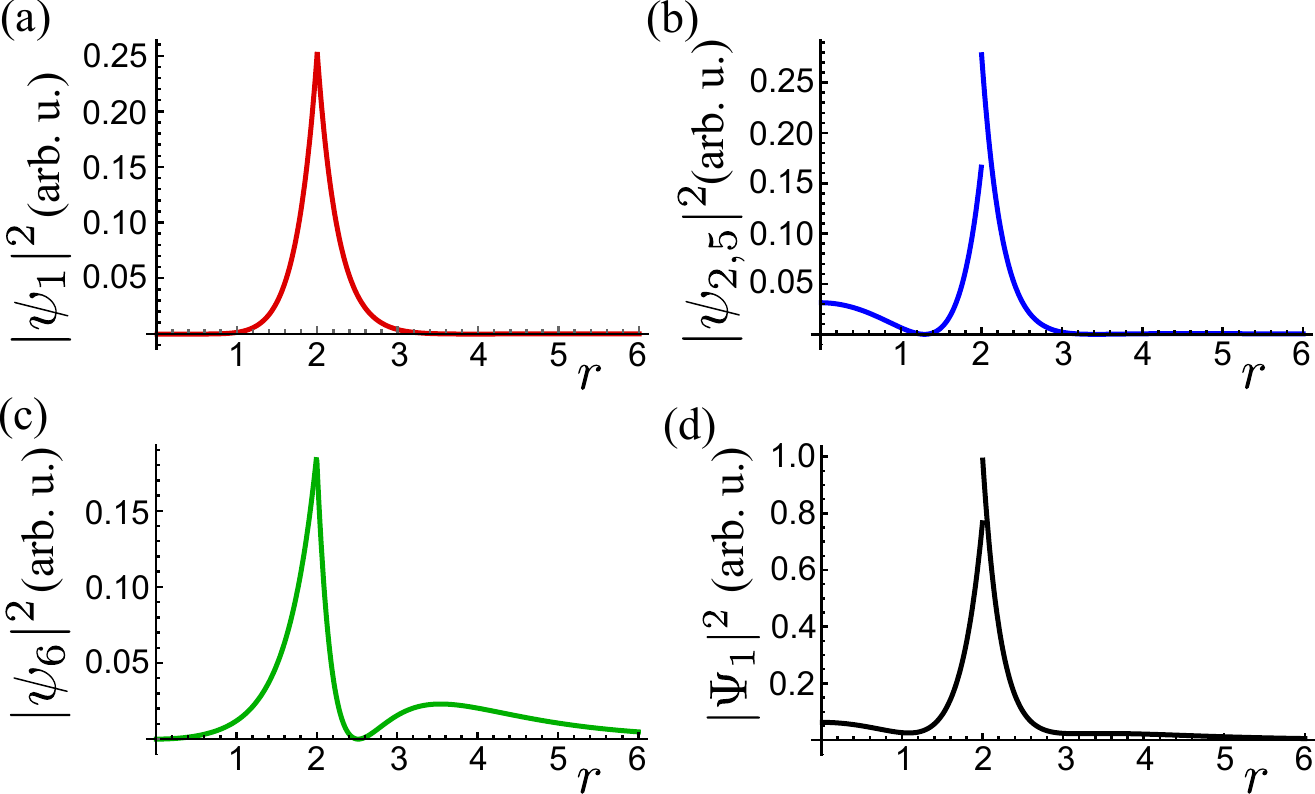}}
\caption{(Color online) The radial distribution of the spinor component densities in the tripletlike states of the second group. Panels (a)--(d) represent the spinor components in the state shown by the line 2.2 in Fig.~\ref{f-triplet_spectrum} at $\varepsilon/2= 0.8717764349$. Other parameters: $a=2.1$, $v_0=2.0$, $r_0=2.0$, $m=0$.}
\label{f-triplet22_w-func}
\end{figure}

The radial distribution of the spinor-component densities is shown in Figs~\ref{f-triplet1_w-func}--\ref{f-triplet22_w-func}. To compare the results with those for the singletlike states, the parameters $a$, $v_0$ and $r_0$ are chosen the same as in Figs~\ref{f-singlet1_w-func} and \ref{f-singlet2_w-func}.
 
In the case of the first group states, Fig.~\ref{f-triplet1_w-func}, the main feature is that the components representing the configuration where the particles are in the different bands, such as $|E\uparrow H\uparrow\rangle$, strongly increase in the tripletlike states in comparison with the corresponding singletlike state. Another peculiarity is that the density of the components corresponding the configuration in which both particles are in the same band, such as $|E\uparrow E\uparrow\rangle$, turns to zero in the center.

The second group of the bound states contains two states with $m=0$, both states being present at $v_0=2$. First, consider the states with lower energy (see the line 2.1 in Fig.~\ref{f-triplet_spectrum}). The spatial distribution of the spinor-component densities in this state is shown in Fig.~\ref{f-triplet21_w-func}. Of largest value are the components in which the particles are in the different bands. Their density is distributed mainly in the bulk of the effective quantum dot, as it is seen in Fig.~\ref{f-triplet21_w-func}(b). This density distribution strongly differs from that in the case of the singletlike states, where the density in the low-energy branch of the second group is located near the edge of the effective quantum dot, Fig.~\ref{f-singlet2_w-func}.

The higher-energy states of the second group (see the line 2.2 in Fig.~\ref{f-triplet_spectrum}) are in contrast located at the edge of the effective quantum dot, Fig.~\ref{f-triplet22_w-func}(a-d). One should note that in the case of the singletlike states, the higher-energy states of the second group have strongly different distribution of the spinor-component densities. When $v_0$ is not small, the predominant components are those in which the particles are in the same bands.

The tripletlike wave functions are antisymmetrized in the same way as described above, so that the antisymmetric wave function reads
\begin{equation}\label{true_triplet_WF}
\begin{split}
 \Psi_m^{(t\uparrow)}(1,2)&= C\left[\psi_{1m}(r)e^{i\varphi}|E\uparrow E\uparrow\rangle\right.\\
 &\left.+ \psi_{2m}(r)\left(|E\uparrow H\uparrow\rangle-|H\uparrow E\uparrow\rangle\right)\right.\\
 &\left.+ \psi_{6m}(r)e^{-i\varphi}|H\uparrow H\uparrow\rangle\right].
\end{split}
\end{equation}

\section{Bound states in topological phase with nearly flat bands}
\label{invert}
We turn to the question of how the coupling of the electron and hole bands affect the bound states in the topological phase where the bands are inverted. The band coupling is characterized by the parameter $a$. When $|a|>2$, the spectrum and the electronic structure of the bound states are little changed qualitatively with varying $a$. However at $|a|<2$ the situation changes in two aspects. 

First, the single-particle wave functions with the energy in the gap are changed radically since the wave vector becomes complex. So that the wave functions not only decay with the distance but also oscillate. The two-particle wave functions behave similarly, since their wave vectors $Q_{\pm}$, see Eq.~(\ref{Q}), are complex. Therefore additional oscillating components appear in the fundamental solutions, such as Eq.~(\ref{Bessel_base}), that form the bound-state wave function. In this case one can expect the appearance of new solutions. 

The second aspect is that the single-particle spectrum also changes essentially with $a$. The spectrum shape changes from nearly parabolic one at $|a|\gg 1$ to that of a mexican-hat form at $|a|<\sqrt{2}$. Correspondingly, the effective mass of electrons also changes very strongly and even changes its sign. The effective mass near the band boundaries is known to play an important role. It is usually supposed that the two-electron bound state is formed due to a negative single-particle energy dispersion near the top band boundary~\cite{gross1971inverse,MahajanJPhysA2006}. 

In this section, we consider the two-particle bound states in the case where $a=\sqrt{2}$. This case is very interesting for two reasons. First, at $a=\sqrt{2}$ the real and imaginary parts of the wave vectors of the states with the energy in the gap are of the same magnitude. Therefore evanescent states in the gap are described by the wave functions which have an oscillating component. Because of this, one can expect that nontrivial interference effects appear in the presence of a spatially inhomogeneous potential. 

Other reason in that the effective mass goes to $\pm \infty$, respectively, at the bottom of the conduction band and the top of the valence band and does not change the sign with changing the energy in the bands. The single-particle energy dispersion is described by the equation $\varepsilon_{\pm}=\pm\sqrt{1+k^4}$. 

The calculations are carried out in the same way as in Sec.~\ref{singlet}. The two-particle energy in the interaction region and outside of it has the form
\begin{equation}
 \varepsilon=2 v_0\Theta(r_0-r)\pm 2\sqrt{1+k^4}\,. 
\end{equation} 

The characteristic wave vectors $Q_{\pm}$ defined by Eq.~(\ref{Q}) take the following values
\begin{align}
 Q_+ &=\sqrt[4]{\widetilde{\varepsilon}^{\,2}-1}\,,\\
 Q_- &=i\sqrt[4]{\widetilde{\varepsilon}^{\,2}-1}\,
\end{align}  
for $\widetilde{\varepsilon}^{\,2}>1$ and 
\begin{equation}
 Q_{\pm}=e^{\pm i \pi/4}\sqrt[4]{1-\widetilde{\varepsilon}^{\,2}}\,.
\end{equation} 
for $\widetilde{\varepsilon}^{\,2}<1$.

The fundamental solutions $\mathcal{F}_m(Qr)$ of the equation system describing the spinor components in the case of both singletlike and tripletlike states are as follows:\\
(i) For $\widetilde{\varepsilon}^{\,2}>1$, the functions $\mathcal{F}_m(Qr)$ are the Bessel functions of the first and second kinds: $J_m(qr)$, $Y_m(qr)$, $I_m(qr)$, and $K_m(qr)$, where $q=\sqrt[4]{\widetilde{\varepsilon}^{\,2}-1}$.\\
(ii) For $\widetilde{\varepsilon}^2<1$, the fundamental solutions are $\mathcal{V}_m^{\pm}=\mathrm{ber}_m(\gamma r)\pm i \mathrm{bei}_m(\gamma r)$ and $\mathcal{W}_m^{\pm}=\mathrm{ker}_m(\gamma r)\pm i \mathrm{kei}_m(\gamma r)$, where  $\mathrm{ber}_m(z)$, $\mathrm{bei}_m(z)$, $\mathrm{ker}_m(z)$, and $\mathrm{kei}_m(z)$ are the Kelvin functions, and $\gamma=\sqrt[4]{1-\widetilde{\varepsilon}^{\,2}}$.

\begin{figure}
\centerline{\includegraphics[width=1.\linewidth]{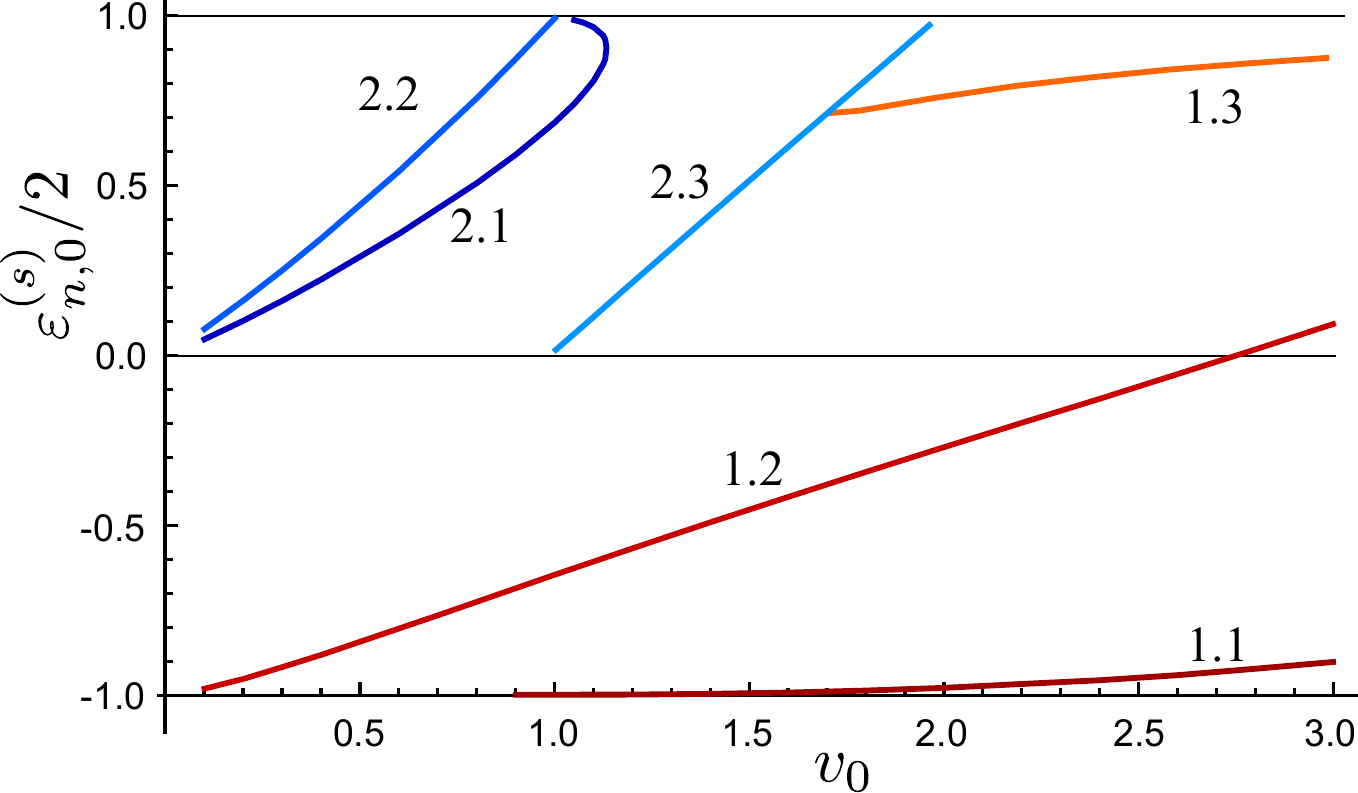}}
\caption{(Color online) The spectrum of the singletlike bound states in TI phase with nearly flat band spectrum ($a=\sqrt{2}$). The  bound-state energy $\varepsilon$ is shown as a function of the interaction potential $v_0$ at $r_0=2.0$ for the states with $m=0$. Lines 1.1, 1.2, 1.3 refer to the states of the first group. Lines 2.1, 2.2 and 2.3 refer to the states of the second group.}
\label{f-singlet_spectrum_1}
\end{figure}

The calculations lead to the following results for the singletlike states. The spectrum of the bound states significantly changes as compared to the case of $a>2$. The main difference is that new states appear. Fig.~\ref{f-singlet_spectrum_1} presents the bound-state spectrum for the same parameters ($r_0$ and $m$) as in Fig.~\ref{f-singlet_spectrum}. New branches are seen to appear in both groups of states.

In the first group, there are two states with $m=0$ at small $v_0$ (the lines 1.1 and 1.2) in contrast to the case of $a>2$ where there is only one state. One of the states has a very small binding energy, while the other state has a large energy and arises at a much smaller interaction potential.

Let us discuss the nature of these states. According to the mechanism of the bound state formation due to the negative single-particle dispersion near the valence-band top~\cite{gross1971inverse}, one could expect that the energy of the bound state will be close to the amplitude of the interaction potential $v_0$, since the effective mass tends to infinity. This really happens with the large-energy state (line 1.2). Obviously, the mechanism of formation of the second state is not directly related to the effective mass in the valence band, since its energy is very close to the bottom of the gap, which indicates an anomalously small effective mass. Therefore we conclude, that the low-energy state (line 1.1) arises because of an interference effect of the evanescent states which results in the appearance of new roots of Eq.~(\ref{DD}). 

\begin{figure}
\centerline{\includegraphics[width=1.\linewidth]{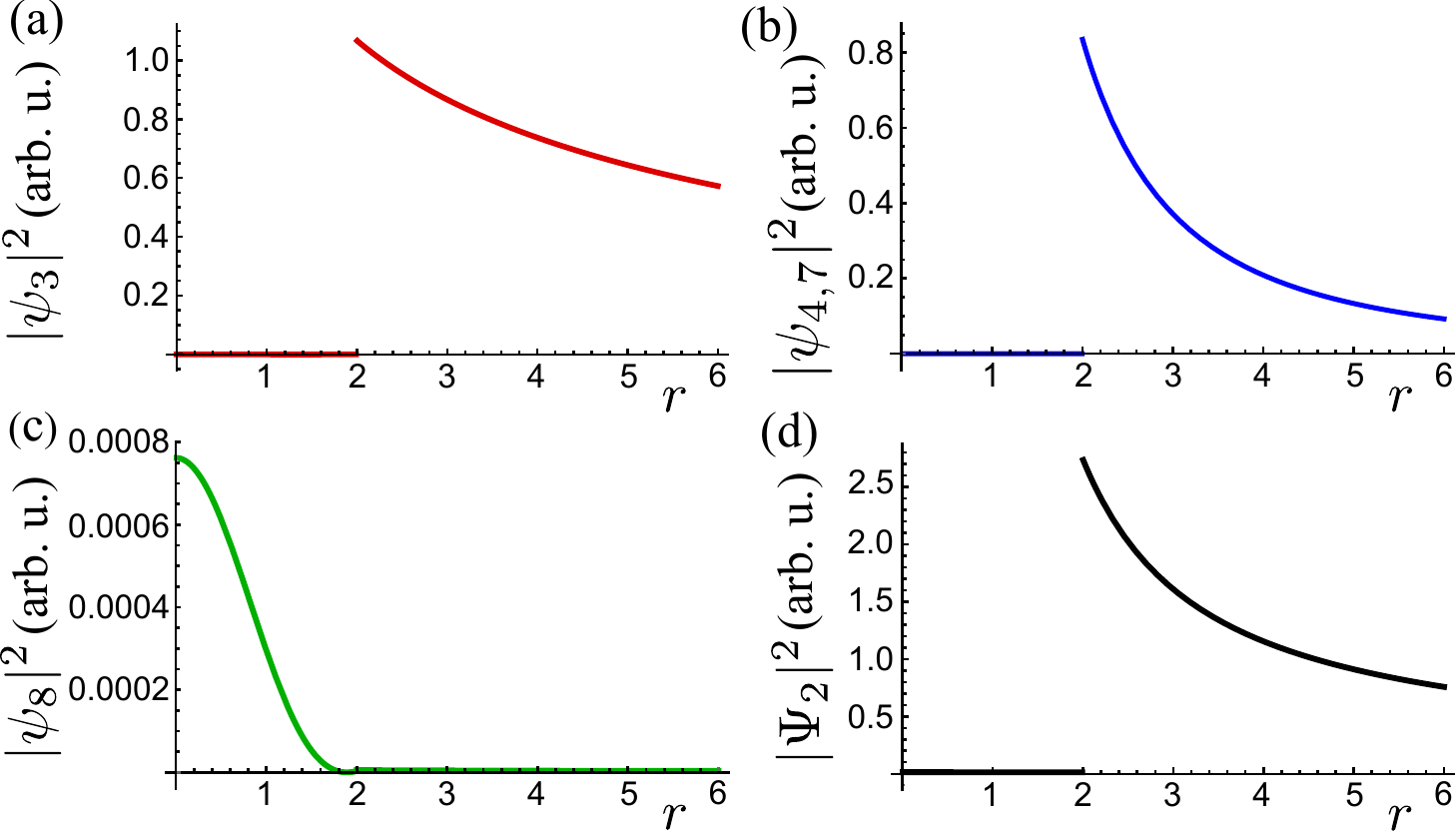}}
\caption{(Color online) The radial distribution of the spinor component densities in the singletlike state shown by the line 1.1 in Fig.~\ref{f-singlet_spectrum_1}.  Panels (a)--(d) represent the spinor components at $v_0=1.0$ and $\varepsilon/2=-0.999954418809$.}
\label{f-singlet1_11_w-func}
\end{figure}

\begin{figure}
\centerline{\includegraphics[width=1.\linewidth]{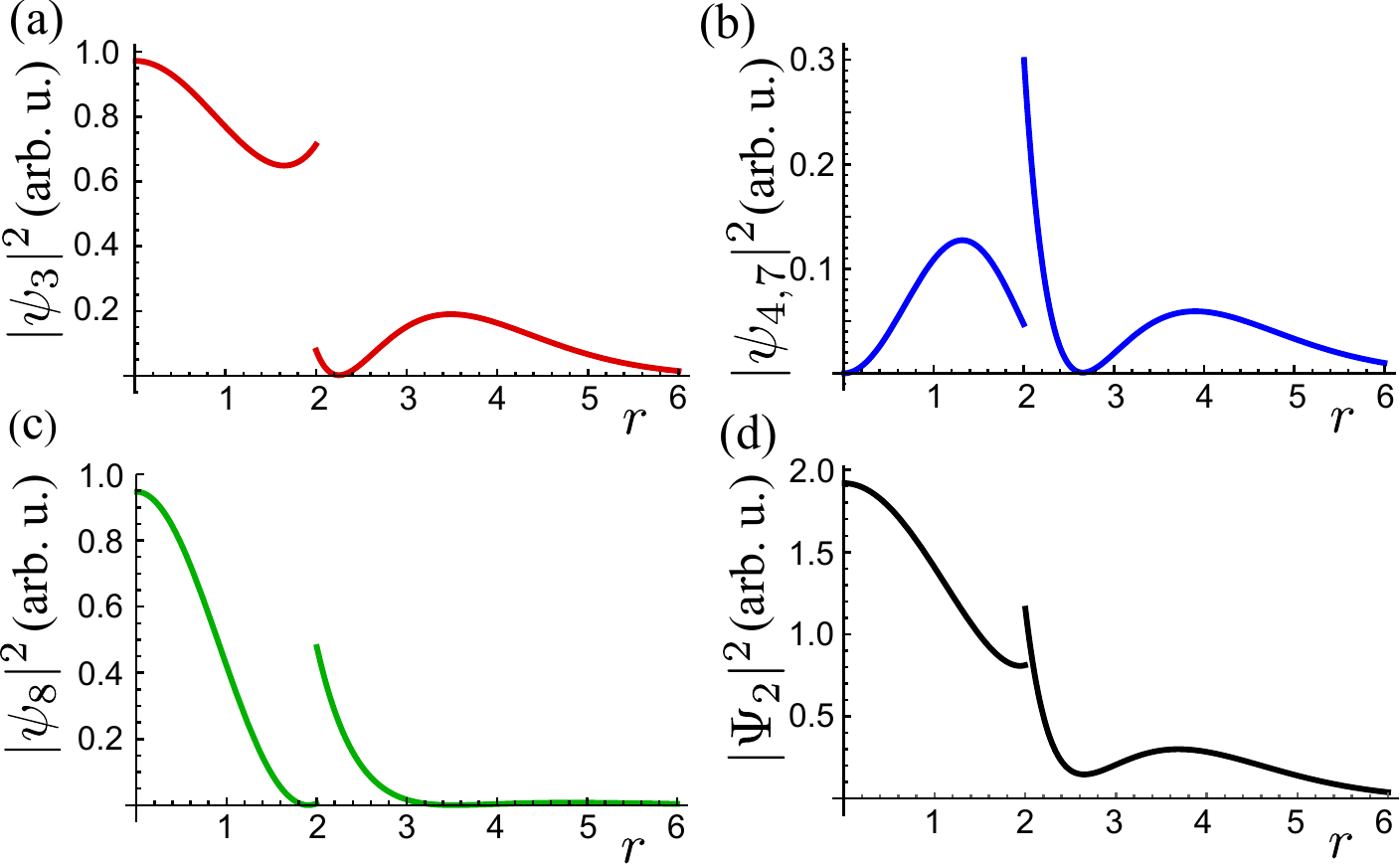}}
\caption{(Color online) The radial distribution of the spinor component densities in the singletlike state shown by the line 1.2 in Fig.~\ref{f-singlet_spectrum_1}. Panels (a)--(d) represent the spinor components at $v_0=1.0$ and $\varepsilon/2=-0.645544446$.}
\label{f-singlet1_12_w-func}
\end{figure}

The states 1.1 and 1.2 differ greatly in the spatial distribution of the density of the spinor components and the magnitude of different components. This is illustrated in Figs~\ref{f-singlet1_11_w-func} and~\ref{f-singlet1_12_w-func}. It can be seen that, in state 1.1, the electron density is mainly localized around the interaction region, which indicates the similarity of this state with the edge state. In state 1.2, on the contrary, the electron density is localized mainly in the effective quantum dot and goes slightly beyond its limits.

Another new state (line 1.3 in Fig.~\ref{f-singlet_spectrum_1}) can not be strictly attributed to any group since it arises at large interaction potential, when $v_0>1.725$. We conventionally classify it to the first group since its energy is lower than $2v_0$. 

The spectrum of the bound states of the second group is also changed compared with the case of $a>2$. Two states (branches 2.1 and 2.2 in Fig.~\ref{f-singlet_spectrum_1}) slightly increase their energy at small $v_0$. With increasing $v_0$, the energy of the state 2.1 becomes a non-monotonic function of $v_0$ as it is shown in Fig.~\ref{f-singlet_spectrum_1}. In addition, a new branch (line 2.3) appears when $v_0>1$ with the energy close to $2v_0$.

\section{Bound states in the topologically trivial phase: band inversion effect}
\label{trivial}
To complete the picture of the two-particle bound states in the BHZ model we present here the results of the study in the case of a topologically trivial phase. In this case, the parameter $\lambda$ in Eqs~(\ref{singlet_gen}), (\ref{triplet_gen}) should be set  equal to $\lambda=+1$ and the calculations are carried out similarly to those described in Secs~\ref{general},~\ref{singlet},~\ref{triplet}. When $\lambda=1$, the fundamental solutions $\mathcal{F}_m(Qr)$ that define the components of the spinors Eq.~(\ref{4_Psi}) are expressed via the Bessel functions of real arguments for any value of the band coupling parameter $a$. Therefore the results do not dramatically depend on $a$. The main result is that two-electron bound states can also exist in topologically trivial phase. Their spectra are not very different from those in the topological phase at $a>2$, but the composition of the spinor components is very different in some cases. Below we demonstrate this for the singletlike states.

\begin{figure}
\centerline{\includegraphics[width=1.\linewidth]{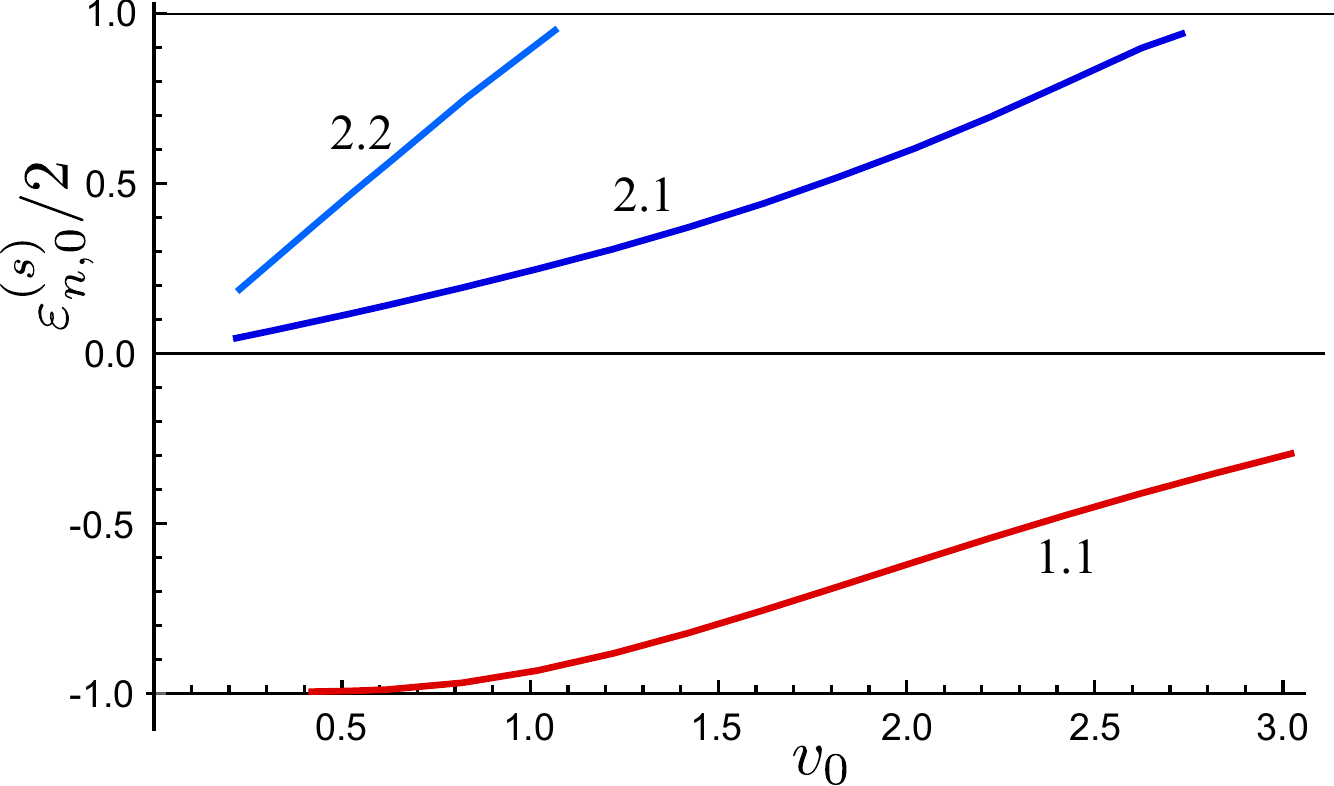}}
\caption{(Color online)  The spectrum of the singletlike bound states in the topologically trivial phase. The bound-state energy $\varepsilon$ is shown as a function of the interaction potential $v_0$ for $a=2.1$, $r_0=2.0$, $m=0$. Line 1.1 shows the the first group states, lines 2.1 and 2.2 refer to the states of the second group.}
\label{f-singlet_spectrum_trivial}
\end{figure}

The spectrum of the singletlike states is presented in Fig.~\ref{f-singlet_spectrum_trivial} for the same parameters ($a=2.1$, $r_0=2.0$ and $m=0$)  as in the case of the topological phase in Fig.~\ref{f-singlet_spectrum}. It is seen that the spectrum of the bound states and the dependence of their energy on the potential amplitude in both cases are qualitatively similar. The energy of the first group state (line 1.1) is somewhat larger than that in the topological phase (line 1.1 in Fig.~\ref{f-singlet_spectrum}). The energies of the states of the second group (lines 2.1 and 2.2) also differ not strongly.

\begin{figure}
\centerline{\includegraphics[width=1.\linewidth]{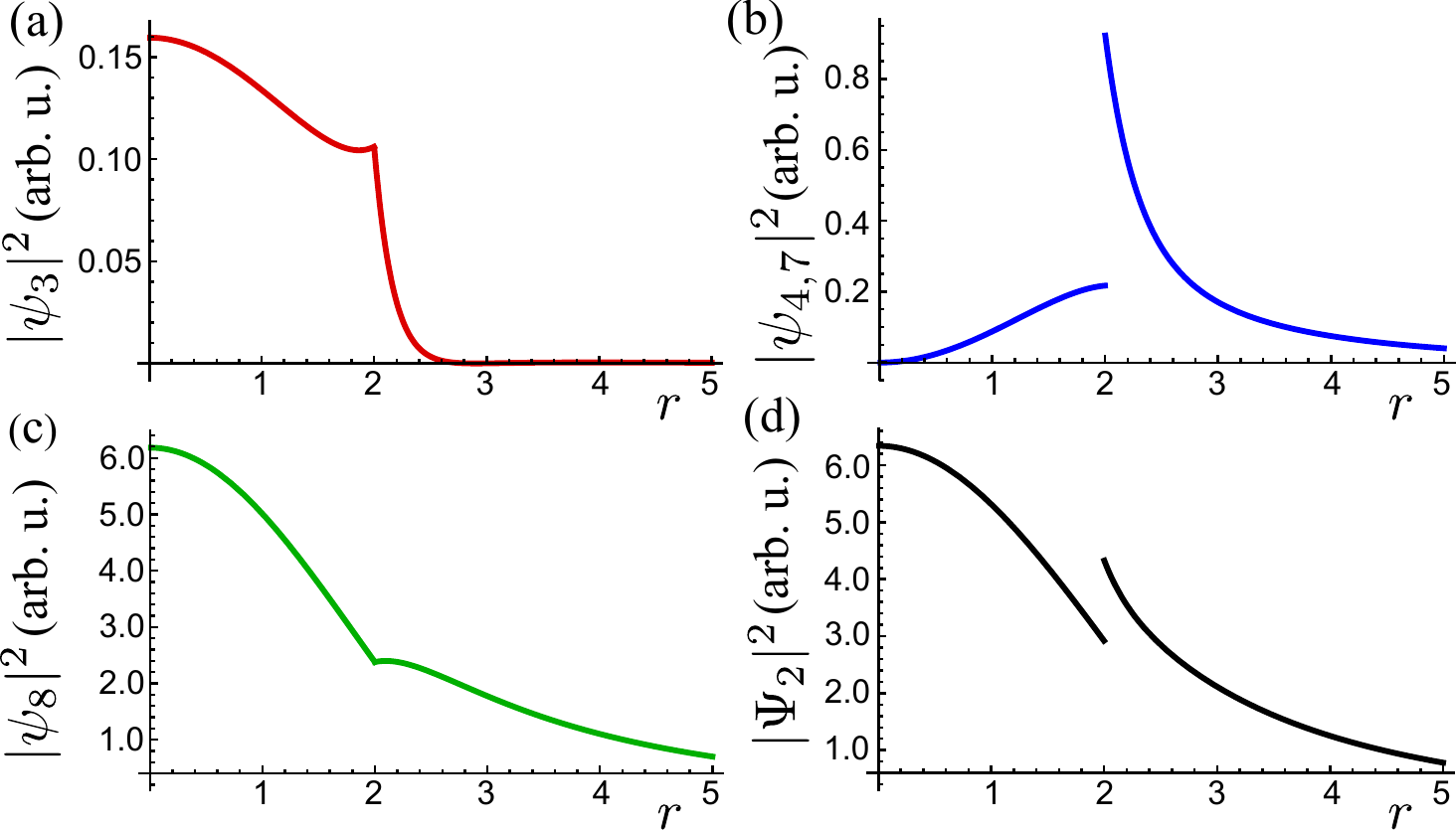}}
\caption{(Color online) The radial distribution of the spinor component densities in the singletlike state shown by the line 1.1 in Fig.~\ref{f-singlet_spectrum_1}. Panels (a)-(d) represent the spinor components at $v_0=1.0$ and $\varepsilon/2=-0.93829486685$.}
\label{f-singlet_triv_11_w-func}
\end{figure}

The spatial distribution of the spinor component densities in the state of the first group is shown in Fig.~\ref{f-singlet_triv_11_w-func}. The main difference from the topologically nontrivial case is that the spinor component $\psi_8$ greatly predominates over the others. 

\begin{figure}
\centerline{\includegraphics[width=1.\linewidth]{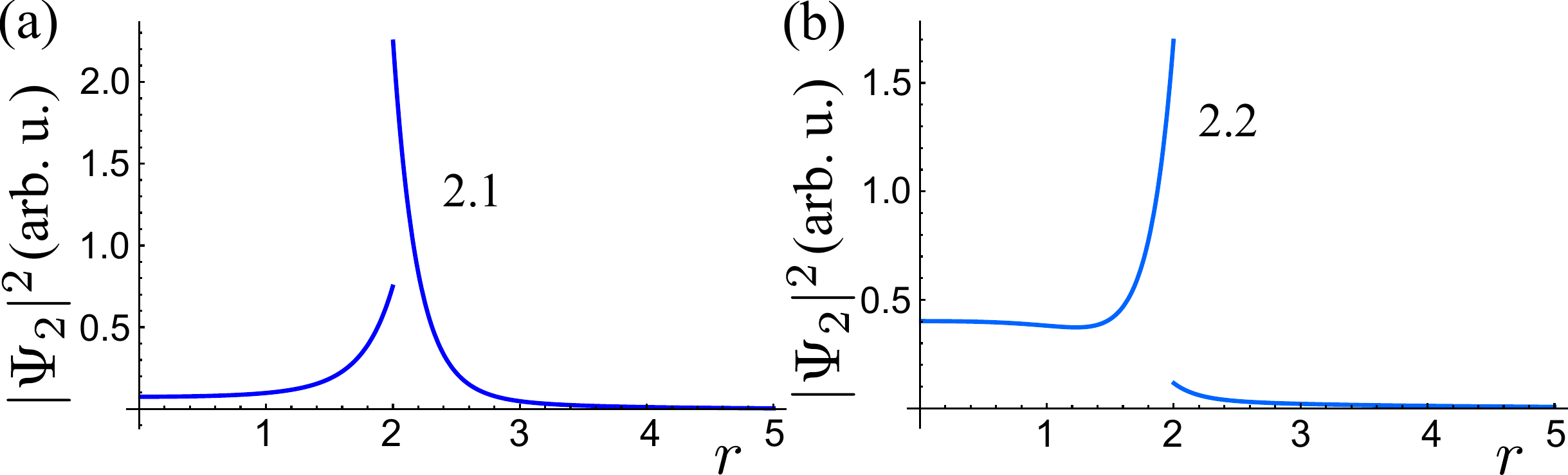}}
\caption{(Color online) The radial distribution of the electron densities in the singletlike states shown by the lines 2.1 (panel (a)) and 2.2 (panel (b)) in Fig.~\ref{f-singlet_spectrum_1} at $v_0=1.0$ and $\varepsilon/2=0.2497827475$ (for the state 2.1) and  $\varepsilon/2=0.922233141$ (for the state 2.2).}
\label{f-singlet_triv_2_w-func}
\end{figure}

The states of the second group do not so strongly differ in the ratio of the spinor components from the topological case. In Fig.~\ref{f-singlet_triv_2_w-func} we present only the spatial distribution of the total electron density for both branches (2.1 and 2.2) of the spectrum.

\subsection{Band inversion effect}
Here we discuss why the energies of the bound states of the first group in the trivial and topological phases are different and what role the band inversion plays in their formation. 

Let us turn again to the mechanism which is commonly used to explain the formation of two-electron bound states in a periodic potential. Two-electron bound states were first discovered in connection with inverse hydrogen absorption spectra observed experimentally~\cite{gross1971inverse}. From the very beginning it was supposed that the bound states are formed due to a negative single-particle energy dispersion near the top band boundary~\cite{gross1971inverse}. This idea was developed within several simple models of periodic potential for one-dimensional~\cite{MahajanJPhysA2006} and 2D systems~\cite{SouzaClaroPRB2010,HaiJPhysCM2014}. This mechanism was used to explain the pairing of repulsive ultracold rubidium atoms in an optical lattice~\cite{winkler2006repulsively} and the observations of two-electron peaks in the photo-double-ionization spectra of aromatic hydrocarbons~\cite{HuberPRA2014}. According to this mechanism the binding energy is determined by the effective mass near the top band boundary of the single-particle spectrum.

It is evident that the results of our calculations do not agree with this idea at least in the case of a topologically nontrivial phase where electron and hole bands are inverted. Indeed, according to the BHZ model the effective mass close to the top of the valence top in the topological phase is larger than in the trivial one. When $|a|>\sqrt{2}$, the ratio of effective masses in topological and trivial phases is equal to $m_{top}/m_{triv}=(a^2+2)/(a^2-2)$, which equals $\approx$2.66 at $a$=2.1. Therefore, it can be expected that the energies of bound states of the first group in the trivial and topological phases will differ significantly. On the contrary, the calculations show that these energies are unexpectedly close (cf.\ lines 1.1 in Figs.~\ref{f-singlet_spectrum} and~\ref{f-singlet_spectrum_trivial}). We argue below that this contradiction indicates that the reduced effective mass, which determines the energy of the bound state, differs significantly from the reduced effective mass, which is determined by the single-particle dispersion, especially in the topological phase where the electron and hole bands are inverted.

First, consider the topologically trivial case. Figure~\ref{f-singlet_triv_11_w-func} shows that the bound state is mainly formed by the single-particle states of the hole band, $|H\uparrow,H\downarrow\rangle$. This is consistent with the structure of the band states near the valence band top. It is well known that in the trivial case these states are formed also by the hole band states. Hence our calculations confirm the known point of view that the two-electron bound states are formed by the single-particle states near the top of the valence band.

Now let us turn to the topologically nontrivial phase. From Fig.~\ref{f-singlet1_w-func} it is seen that the bound state is mainly formed by the basis states of both the electron and hole bands, $|E\uparrow,E\downarrow\rangle$ and $|H\uparrow,H\downarrow\rangle$. Furthermore, both bands contribute to the total density almost equally. Direct calculations for a variety of the interaction potential amplitudes (not shown here) confirm this result even in if the binding energy is very small. This is easy to understand, since the energy of the bound state lies in the gap where the electron and hole bands overlap, and therefore the electron-band states largely contribute to the total wave function. In contrast, in the trivial case the bands do not overlap. 

It should be noted that the spinor structure of the bound state is not consistent with that of the band states at the top of the valence band. In the case of inverted bands, the valence band states are well known to be formed by the electron band states. This is easy to see directly from the Hamiltonian~(\ref{h(k)}) at $a>\sqrt{2}$ and $k\to 0$. 

Since the spinor structures of the bound states and valence band states are very different, there are no arguments to think that the reduced effective masses are the same in both cases. Because the bound states are formed by a mixture of the states of the electron and hole bands and their products (specifically, $|E\uparrow,E\downarrow\rangle$, $|E\uparrow,H\downarrow\rangle$, $|H\uparrow,E\downarrow\rangle$, and $|H\uparrow,H\downarrow\rangle$), the reduced effective mass depends on the weights of the components, which should be found by a direct solution of the two-particle Schr\"odinger equation with a given interaction potential. It is evident that these weights are dependent on the profile and amplitude of the interaction potential.

\section{Conclusions}
\label{conclude}
In this paper we have investigated the two-body problem for 2D electron system described by the symmetric BHZ model in the case of both topologically nontrivial and trivial phases. The main conclusion is that the interaction between the electrons leads to the formation of two-particle bound states at any sign of the pair interaction potential. The pairing of electrons under the action of a repulsive potential becomes possible due to the formation of a negative reduced effective mass of two electrons. The two-electron bound states have the charge $2e$ like the Cooper pair, but their energy lies in the gap of the band spectrum. In this respect they are akin to excitons.

In the case where the spin $S_z$ in conserved in the single-particle states, the two-particle states are classified according to the moments of their constituent electrons as a singletlike state (with opposed moments of the electrons) and two tripletlike states (with parallel electron moments).

The bound state spectrum has been studied in a simplified case of zero total momentum of the pair. In general, the spectrum is dependent on the total momentum since the relative motion of the electrons is coupled to the motion of the center of mass. The bound state energy lies in the gap of the two-electron band spectrum. Since the interaction potential is a function only of the distance between electrons, the bound states are specified by an angular quantum number $m$. General properties of the bound state spectrum are found with using a steplike model potential. The states are well classified into two groups in the case of small interaction potential amplitude when they noticeably differ in the energy. The states of the first group have the energy close to the bottom of the gap of the band spectrum, and the energy of the second group states lies near the middle of the gap.
 
In the trivial phase, the states of the first group are mainly formed by the basis states in which both electrons are mainly in the hole-band states, such as $|H\uparrow,H\downarrow\rangle$. In this case, it is obvious that the reduced effective mass is negative. At a given angular quantum number the bound states are specified by the radial quantum number, so that there is a series of these states. The states of the second group are mainly formed by the basis states composed of different bands such as $|E\uparrow,H\downarrow\rangle$. In these states, one of the electrons is the electron band and the other is in the hole band. It is clear that in this case the reduced effective mass also can be negative. Interestingly, in this group of states there are two states at a given $m$ and no other states appear with increasing the interaction radius $r_0$ in the range we have studied, though the bound state energies depend on $r_0$. Only one of the state can disappear with increasing $r_0$.

In the topologically nontrivial phase, the situation is more complicated. It depends on the coupling of the electron and hole bands. When the coupling parameter is large, $|a|>2$, the bound state spectrum at small $v_0$ is qualitatively similar to that in the trivial case, though the energies are noticeably changed. However, the electronic structure of the first group states is changed dramatically. In contrast to the trivial case, these states are formed by the basis states of both bands, to be exact, by the states $|E\uparrow,E\downarrow\rangle$ and $|H\downarrow,H\uparrow\rangle$. That is the bound states are a superposition of the states in which both electrons are in the hole and electron band states. Due to this fact the binding energy turns out to substantially increase as compared with the trivial case. 

When the coupling of the bands is not strong, $|a|<2$, the situation changes radically as a consequence of the fact that the evanescent states forming the bound state in the gap contain an oscillating component. We demonstrate this by a detailed study of the case where $a=\sqrt{2}$, which allows one to find the solution exactly. In this case the single-particle spectrum is nearly flat at the band boundaries. The bound state spectra are strongly changed in both groups. The main effect is that new bound states arise in the spectrum in addition to the states of the same type as in the case of $|a|>2$. Of particular interest is the new state appearing in the first group. The new state has a much higher binding energy and arises at much lower interaction potential than other states. This fact shows that the band inversion can favor pairing the electrons when the band coupling is not strong.

The mechanism of the band-inversion impact on the formation of two-particle bound states is caused by two factors:  a strong change in the composition of the basis states, which mainly form a given bound state because of the band inversion, and the appearance of oscillating evanescent states.

In the trivial phase, the states of the first group are formed mainly by the basis states of the hole band, such as $|H\uparrow, H\downarrow\rangle$. In contract, in the inverted-band case with strong coupling, $|a|>2$, the bound states of the first group are mainly formed by the basis states of both the electron and hole bands, such as $|H\uparrow, H\downarrow\rangle$ and $|E\uparrow, E\downarrow\rangle$, even if the interaction potential is small. 

In the case of nearly flat bands, $|a|=\sqrt{2}$, new addition states arise in the bound state spectrum. In the new state, the weight of the mixed basis states, such as $|E\uparrow, H\downarrow\rangle$, is noticeably increased. Since the bound states are formed by the mixture of the two-particle basis states, which strongly differs from that forming the conduction and valence bands, the reduced effective mass, which appears in the bound state formation, can be essentially different from the reduced effective mass determined by the band spectrum. The weights of the basis states in the bound state are determined by the solution of the Schr\"odinger equations, like Eqs.~(\ref{triplet_gen}) and (\ref{singlet_gen}), with a given interaction potential. Unfortunately, we failed to find any general relationships for the reduced effective mass in the two-particle bound state. 

To complete the picture it is interesting to study the case when the single-particle spectrum in the bands is of a mexican-hat shape. However, this situation  requires a separate careful study because of strong singularity of density of states, which appears in this case at the boundaries of the single-particle band spectrum. It is well known that such singularity remarkably facilitates the pairing of electrons~\cite{ChaplikPRL2006,CappellutiPRL2007,TakeiPRA2012,he2014generic,GoldsteinPRB2015}.

Another interesting point refers to the life time of the two-electron bound states. Since the bound states have the energy in the band gap, they can decay into the states of noninteracting electrons under the action of external disturbances. However, the probability of this decay is greatly reduced if the lower band is filled by electrons. In this connection an important question arises about the life time of the bound states in the presence of many electrons. This question also requires a separate study. 

\acknowledgments
This work was supported by Russian Science Foundation under the grant No.~16-12-10335 in the part related to the general properties of two-particle states (Sec.~\ref{general}) and numerical calculations, and by Russian Foundation for Basic Research under the grant No.~17-02-00309.

\bibliography{two-body_BHZ}
\end{document}